%% file: main.tex
\documentclass[12pt,a4paper]{article}
\usepackage[english]{babel}
\usepackage[usenames, dvipsnames]{color}
\usepackage{jheppub}
\usepackage{amsmath}  
\usepackage{tabularx} 
\usepackage{graphicx} 
\usepackage{float}
\usepackage{amssymb}
\usepackage{bm}
\usepackage{tikz}
\usepackage{tikz-feynman}
\tikzfeynmanset{compat=1.1.0}
\usetikzlibrary{shapes,arrows,positioning,automata,backgrounds,calc,er,patterns}
\usepackage{mathtools}
\usepackage{wasysym}
\usepackage{ tipa }
\usepackage{ marvosym }
\usepackage{ bbold }
\usepackage{amstext} 
\usepackage{array}
\usepackage{slashed}
\usepackage{physics}
\usepackage{ctable}
\usepackage{blindtext}
\usepackage[T1]{fontenc}
\newcolumntype{L}{>{$}l<{$}}
\hypersetup{
	colorlinks=true,       
	linkcolor=blue,        
	citecolor=blue,        
	filecolor=magenta,     
	urlcolor=blue
}

\newcommand{\Op}{\mathcal{O}}

\newcommand{\lvector}[2]{\begin{pmatrix}
#1\\
#2
\end{pmatrix}}

\newcommand{\iM}{i\mathcal{M}}

\newcommand{\pl}{\left(1 - \gamma_5\right)}

\newcommand{\lr}[1]{\left(#1\right)}

\newcommand\brobor{\smash[b]{\raisebox{0.6\height}{\scalebox{0.5}{\tiny(}}{\mkern-1.5mu\scriptstyle-\mkern-1.5mu}\raisebox{0.6\height}{\scalebox{0.5}{\tiny)}}}}

\title{Constraining lepton number violating interactions in rare kaon decays}

\preprint{TUM-HEP-1274/20}

\author[a]{Frank F. Deppisch}
\author[b]{K\aa{}re Fridell}
\author[b]{Julia Harz}

\emailAdd{f.deppisch@ucl.ac.uk, kare.fridell@tum.de, julia.harz@tum.de}
\affiliation[a]{Department of Physics and Astronomy, University College London, Gower Street,\\ London WC1E 6BT, United Kingdom}
\affiliation[b]{Physik Department T70, James-Franck-Stra{\ss}e, Technische Universit\"at  M\"unchen,\\ 85748 Garching, Germany}

\makeatletter
\def\@fpheader{\relax}
\makeatother

\date{\today}

\abstract{
\noindent We investigate the possibility to probe lepton number violating (LNV) operators in the rare kaon decay $K\to\pi\nu\nu$. Performing the analysis in the Standard Model effective field theory with only light active Majorana neutrinos, we determine the current limits on the corresponding LNV physics scale from the past E949 experiment at BNL as well as the currently operating experiments NA62 at CERN and KOTO at J-PARC. We focus on the specific signature of scalar currents in $K\to\pi\nu\nu$ arising from the LNV nature of the operators and study the effect on the experimental sensitivity, stressing the need for dedicated searches for beyond the SM currents. We find that the rare kaon decays probe high operator scales $\Lambda_\text{LNV} \approx 15$ to 20~TeV in different quark and neutrino flavours compared to neutrinoless double beta decay. Furthermore, we comment that the observation of LNV in kaon decays can put high-scale leptogenesis under tension. Finally, we discuss the connection with small radiatively generated neutrino masses and show how the severe constraints therefrom can be evaded in a minimal ultraviolet-complete scenario featuring leptoquarks.}

\begin{document}
\maketitle

\input{intro.tex}
\input{operators.tex}
\input{rarekaon.tex}
\input{leptoquarks.tex}
\input{conclusions.tex}

\bibliographystyle{JHEP}
\bibliography{References}

 \end{document}

%% file: intro.tex
\section{Introduction}

The nature of neutrinos and especially the origin of their masses are a crucial open question in particle physics. The Standard Model (SM) incorporates a successful and experimentally verified mechanism to give mass to the charged fermions despite the fact that bare mass terms are forbidden due to the chiral nature of the SM. The latter is especially apparent for neutrinos: while the so-called active neutrinos $\nu_L$ form $SU(2)_L$-doublets with the left-handed charged leptons, the corresponding right-handed $SU(2)_L$-singlet neutrino states $\nu_R$ that would be needed to give neutrinos a so-called Dirac mass, would also be required to be uncharged under the SM hypercharge $U(1)_Y$. Hence, they would be completely \emph{sterile} under the SM gauge interactions. Out of this reason they have been omitted in the SM for economical purposes, however, from neutrino oscillations \cite{Agashe:2014kda} we know that at least two out of the three known neutrino species have finite masses. Oscillations themselves are only sensitive to neutrino mass-squared differences, pointing to mass splittings of the order $10^{-2}$~eV to $5\times 10^{-2}$~eV. Combining this knowledge with the most stringent upper limits on absolute neutrino masses from Tritium decay \cite{Aker:2019uuj} and cosmological observations \cite{Ade:2015xua} finally constrain all neutrinos to be lighter than $m_\nu \approx 0.1$~eV.

Incorporating Dirac neutrino masses via the SM Higgs mechanism is generally possible, but would lead to two theoretical issues: (i) the Yukawa couplings with the Higgs are tiny, $y_\nu \sim m_\nu / \Lambda_\text{EW} \lesssim \mathcal{O}(10^{-12})$, with the electroweak (EW) scale $\Lambda_\text{EW}$ and (ii) total lepton number $L$ is no longer an accidental symmetry due to the required presence of the sterile right-handed neutrinos. Specifically, the right-handed neutrinos are allowed to have a \emph{Majorana} mass $M$ of the form $M \bar \nu_R C \bar\nu_R^T$ violating total lepton number by two units, $\Delta L = 2$, due to the $\nu_R$ being sterile under the SM gauge group and thus unprotected by its otherwise chiral nature. If lepton number is not explicitly conserved, neutrinos are expected to be of Majorana nature. The most prominent realisation of Majorana neutrino masses is the Seesaw mechanism, in which the $\nu_R$ acquires a large Majorana mass term, $M \gg v$, where $v$ is the SM Higgs vacuum expectation value (VEV), which then mediates to the active neutrino $\nu_L$ via the Yukawa couplings \cite{Minkowski:1977sc, Mohapatra:1979ia, Yanagida:1979as, seesaw:1979, Schechter:1980gr}. Within this mechanism, a high scale $M \approx 10^{14}$~GeV can naturally explain the light neutrino masses $m_\nu \approx 0.1$~eV.

However, the most prominent scenario, the high-scale seesaw mechanism, is not the only way to generate light Majorana neutrino masses. Other possibilities include incorporating lepton number violation (LNV) at low scales in secluded sectors, at higher loop order or allowing for higher-dimensional effective interactions. If the breaking of the lepton number symmetry occurs close to the EW scale, higher-dimensional $L$-breaking operators can be important. Thus, from a phenomenological point of view, searching for processes that violate total lepton number plays a crucial role in neutrino and Beyond-the-SM (BSM) physics.

Lepton number violation may also be relevant for the generation of matter in the universe. In Leptogenesis scenarios within the context of seesaw mechanisms for neutrino mass generation, lepton number violating processes are required to occur out of equilibrium in the early Universe. This also implies that LNV interactions cannot be too large, otherwise the resulting processes will be washing out a lepton number asymmetry before it can be transformed to a baryon asymmetry via SM sphaleron processes. In a framework of LNV operators, this would allow to set upper limits on the scale of leptogenesis (or baryogenesis in general) if any LNV process is observed \cite{Deppisch:2013jxa, Deppisch:2015yqa, Deppisch:2018}.

In the context of Majorana neutrino masses and LNV in general, the search for neutrinoless double beta ($0\nu\beta\beta$) decay is considered to be the most sensitive possibility to probe Majorana neutrino masses. The experimentally most stringent lower limit on the decay half life $T_{1/2}$ is derived using the Xenon isotope ${}^{136}_{\phantom{1}54}$Xe,
\begin{align}
T_{1/2}^\text{Xe} \equiv T_{1/2}\left({}^{136}_{\phantom{1}54}\text{Xe} \to {}^{136}_{\phantom{1}56}\text{Ba} + e^- e^-\right) \gtrsim 10^{26}~\text{y}.
\end{align}
However, Majorana neutrino masses are not the only possibility within BSM physics that can induce $0\nu\beta\beta$ decay. High-scale New Physics (NP) can similarly contribute to effective low-energy operators leading to $0\nu\beta\beta$ decay \cite{Pas:1999fc, Pas:2000vn, delAguila:2011gr, delAguila:2012nu}. Hereby, one assumes that there are no other exotic BSM particles below the $0\nu\beta\beta$ energy scale of $m_F \approx 100$~MeV. The standard mass contribution (Weinberg-operator) usually considered corresponds to a so-called long-range transition via the exchange of a light neutrino. Here, the $0\nu\beta\beta$ decay rate can be estimated on dimensional grounds as $\Gamma^{0\nu\beta\beta}_{m_\nu} \sim m_\nu^2 G_F^4 m_F^2 Q_{\beta\beta}^5 \sim (m_\nu/0.1~\text{eV})^2 (10^{26}~\text{y})^{-1}$. $G_F$ indicates the SM Fermi coupling and the phase space available to the two electrons scales as $Q_{\beta\beta}^5$ with $Q_{\beta\beta} = \mathcal{O}(1~\text{MeV})$ for typical double beta decay nuclear transitions. In models with exotic interactions, no mass insertion is required. Instead, the decay rate can be estimated as $\Gamma^{0\nu\beta\beta}_\text{LR} \sim v^2 \Lambda_{O^{(7)}}^{-6} G_F^2 m_F^4 Q_{\beta\beta}^5 \sim (10^5~\text{GeV}/\Lambda_{O^{(7)}})^6 (10^{26}~\text{y})^{-1}$, with the SM Higgs VEV $v$ and the scale $\Lambda_{O^{(7)}}$ of the dim-7 operator. Such exotic long-range mechanisms have received considerable attention so far, see e.g. \cite{Doi:1981}. This is understandable as the suppression at dimension-7 is still fairly low while $0\nu\beta\beta$ decay is sensitive to high scales of order $\Lambda_{O^{(7)}} \approx 10^5$~GeV.

$0\nu\beta\beta$ decay can only probe LNV interactions among first generation fermions. Instead, we will focus on meson decays which are recognized as important probes of exotic physics \cite{Buras:2013ooa, Isidori:2014rba, Buras:2014zga}. NP contributions to meson decays are expected to occur at a high energy scale and can be model-independently described by effective operators. In the context of LNV, kaon decay modes such as $K^+ \to \pi^-\ell^+\ell^+$ \cite{Littenberg:1991ek,Littenberg:2000fg,Chun:2019nwi} are explicitly violating charged lepton number by two units and their discovery would establish that neutrinos are Majorana fermions. On the other hand, in the decay modes $K^+ \to \pi^+ \nu\bar\nu$ and $K_L \to \pi^0 \nu\bar\nu$, the emitted neutrinos are not observed. While in the SM with conserved lepton number one $\nu$ is considered to be a neutrino and the other an anti-neutrino, if neutrinos are Majorana fermions, the process can also be interpreted as LNV with the emission of two neutrinos or two anti-neutrinos. Under the absence of sterile neutrinos, this requires the participation of the right-chiral neutrino component (a Majorana neutrino is constructed as $\nu = \nu_L + \nu_L^c$). As we will demonstrate, the corresponding change in the helicity structure is in principle observable and can be used to distinguish the lepton number conserving (LNC) and LNV modes.

Within the SM with only left-handed neutrinos, the rare kaon decay of the form $K\to \pi \nu\bar\nu$ can be effectively described by the dimension-6 operator of the form $d^c s_L \nu_L^c \nu_L$ (in terms of the $SU(2)_L$ component fields). In the SM, it is generated through loops involving $W$ and $Z$ bosons, see Fig.~\ref{fig:SMdiagrams}. Due to GIM suppression and loop suppression, the branching ratio of the decays are very small, BR$(K\to \pi\nu\bar\nu)\approx (3-9)\times 10^{-11}$. This decay is thus very sensitive to exotic effects and NP contributions at scales of order $\Lambda \approx 200$~TeV can be probed \cite{Buras:2014zga}. At dimension-6 no SM-invariant operators that violate lepton number by two units exist. Instead, the lowest dimension at which a LNV operator can lead to a short-range contribution in the rare kaon decay $K\to \pi \nu\bar\nu$ is at dimension-7 \cite{Babu:2001}. The operator is of the form $h_0 d^c s_L \nu_L \nu_L$ and its contribution to the rare kaon decay is illustrated in Fig.~\ref{fig:Op3bdiagram}~(left), with the Higgs field involved acquiring its VEV. We will discuss this scenario and operator contribution in detail to determine the sensitivity of current and planned searches for the kaon decay mode and how it differs from the usually considered LNC case.

The paper is organized as follows. In Section~\ref{sec:operators} we discuss the effective LNV operators relevant for our analysis and their consequences in $0\nu\beta\beta$ decay, radiative neutrino mass corrections and leptogenesis. Section~\ref{sec:rarekaon} contains a detailed analysis of rare kaon decays, with a focus on $K\to\pi\nu\nu$, and their relevance in probing LNV operators. This most importantly includes a discussion of the difference to the effect of the usually considered lepton number conserving operators beyond the SM. In Section~\ref{sec:leptoquarks}, we present an ultraviolet complete scenario using leptoquarks to illustrate an example of how the effective LNV operators can be generated in NP. We conclude our discussion in Section~\ref{sec:summary}.

%% file: operators.tex
\section{Lepton Number Violation}
\label{sec:operators}

In the following, we discuss in more detail the possibility to probe LNV interactions with meson decays. We first introduce LNV operators and highlight the most relevant in this context. We will then discuss their connection to rare kaon decays and the possibility to radiatively generate neutrino masses. Finally, we comment on the consequences of observing LNV on the viability of Leptogenesis scenarios.

\subsection{LNV operators}

In order to model independently describe lepton number violating interactions, the SM effective field theory (SMEFT) approach is a powerful tool. SMEFT contains all SM fields and describes NP contributions in the form of Lorentz invariant non-renormalizable operators that are invariant under the SM symmetry group $SU(3)_C\times SU(2)_L\times U(1)_Y$. The NP contributions are absorbed in the corresponding Wilson coefficients or the scale of the new operators. The lowest dimensional operator when extending the SM by LNV interactions, is the well-known Weinberg operator~\cite{Weinberg:1979sa}
\begin{align}
\label{eq:Weinberg}
	\Op^{(5)}_1 = L^\alpha L^\beta H^\rho H^\sigma \epsilon_{\alpha\rho} \epsilon_{\beta\sigma},
\end{align}
at dimension-5. Generally, LNV operators can occur only at odd mass dimensions such that the SM Lagrangian $\mathcal{L}_{SM}$ can be extended as
\begin{align}
\label{eq:OpSeries}
	\mathcal{L} = \mathcal{L}_{SM} + \frac{1}{\Lambda_1}\mathcal{O}_1^{(5)} +
		\sum_i \frac{1}{\Lambda^3_i}\mathcal{O}^{(7)}_i +
		\sum_i \frac{1}{\Lambda^5_i}\mathcal{O}^{(9)}_i + \cdots,
\end{align}
where $\mathcal{O}^{(D)}_i$ indicates the SM effective operators at dimension $D = 5, 7, 9, \dots$ that are correspondingly suppressed by the scale $\Lambda^{D-4}_i$ of NP. The index $i$ labels the individual LNV operators. All LNV operators up to dimension 11 have been identified in the literature~\cite{Babu:2001, deGouvea:2008, Deppisch:2018}, apart from those including gauge bosons or derivatives, as they may be more difficult to incorporate in renormalizable ultraviolet (UV) complete theories at tree level~\cite{Babu:2001}. These are nevertheless interesting, and gauge invariant LNV operators with derivatives could be searched for at the LHC in same-sign gauge boson fusion \cite{Aoki:2020til}. However, for a given LNV process, operators with derivatives generally occur at higher dimension than those without, which is why we neglect such operators here. The operator scales $\Lambda_i$ in Eq.~\eqref{eq:OpSeries} subsume all NP parameters such as coupling strengths and masses of the complete UV theory that is integrated out. This implies at the same time that the SMEFT approach is only valid up to the corresponding scale.

The fields that enter into LNV SMEFT operators are the SM fermion and Higgs fields,
\begin{equation}
\label{eq:SMfields}
    L^{\alpha} = \lvector{\nu_L}{e_L}^{\alpha}, \hspace{5pt} Q^{\alpha} = \lvector{u_L}{d_L}^{\alpha}, \hspace{5pt} H = \lvector{h^+}{h^0}, \hspace{5pt} e^c_{\alpha}, \hspace{5pt} u^c_{\alpha}, \hspace{5pt} d^c_{\alpha}\\,
\end{equation}
where, the superscript $c$ denotes charge conjugation and $\alpha$ indicates the flavour. All fermion fields in Eq.~\eqref{eq:SMfields} are left-handed 2-component Weyl spinors, and $L^\alpha, Q^\alpha, H^\alpha$ are $SU(2)_L$ doublets. For our purposes, we assume in the following that the neutrino is a Majorana particle, i.e. a four-component spinor that is constructed as $\nu=\nu_L + \nu^c_L$. Out of the SM fields we can then construct LNV operators. We will discuss them in more detail in the following section.
\begin{table}
\centering
\def\arraystretch{1.1}
\begin{tabular}{L|L}
\specialrule{.2em}{.3em}{.0em}
i & \mathcal{O}_i \\
\hline
1 & L^\alpha L^\beta H^\rho H^\sigma \epsilon_{\alpha\rho} \epsilon_{\beta\sigma} \\
1^{y_d} & L^\alpha L^\beta H^\rho H^\sigma \bar{Q}^{\eta}H_{\eta}\bar{d}^c \epsilon_{\alpha \rho}\epsilon_{\beta \sigma} \\
3a &  L^\alpha L^ \beta Q^\rho d^cH^\sigma\epsilon_{\alpha \beta}\epsilon_{\rho \sigma} \\
3a^{H^2} &  L^\alpha L^ \beta Q^\rho d^cH^\sigma\bar{H}^{\eta}H_{\eta}\epsilon_{\alpha \beta}\epsilon_{\rho \sigma} \\
3b &  L^\alpha L^ \beta Q^\rho d^cH^\sigma\epsilon_{\alpha \rho}\epsilon_{\beta \sigma} \\
3b^{H^2} &  L^\alpha L^ \beta Q^\rho d^cH^\sigma\bar{H}^{\eta}H_{\eta}\epsilon_{\alpha \rho}\epsilon_{\beta \sigma} \\
4a &  L^\alpha L^ \beta \bar{Q}_\alpha \bar{u}^cH^\rho\epsilon_{\beta \rho} \\
4a^{H^2} &  L^\alpha L^ \beta \bar{Q}_\alpha \bar{u}^cH^\rho\bar{H}^{\sigma}H_{\sigma}\epsilon_{\beta \rho} \\
4b^{\dagger} &  L^\alpha L^ \beta \bar{Q}_\rho \bar{u}^cH^\rho\epsilon_{\alpha \beta} \\
4b^{\dagger H^2} &  L^\alpha L^ \beta \bar{Q}_\rho \bar{u}^cH^\rho\bar{H}^{\sigma}H_{\sigma}\epsilon_{\alpha \beta} \\
5 & L^\alpha L^\beta Q^\rho d^c H^\sigma H^\eta\bar{H}_{\alpha}\epsilon_{\beta\sigma}\epsilon_{\rho\eta} \\
6 & L^\alpha L^\beta \bar{Q}_\rho \bar{u}^cH^\sigma H^\rho\bar{H}_\alpha\epsilon_{\beta\sigma} \\
7 & L^\alpha Q^\beta\bar{e}^c\bar{Q}^\rho H_\rho H^\sigma H^\eta\epsilon_{\alpha\sigma}\epsilon_{\beta\eta} \\
8 & L^\alpha \bar{e}^c\bar{u}^c d^cH^\beta\epsilon_{\alpha \beta} \\
\specialrule{.2em}{.0em}{.3em}
\end{tabular}
\qquad\qquad
\begin{tabular}{L|L}
	\specialrule{.2em}{.3em}{.0em}
	i & \mathcal{O}_i \\
	\hline
	8^{H^2} & L^\alpha \bar{e}^c\bar{u}^c d^cH^\beta\bar{H}^{\rho}H_{\rho}\epsilon_{\alpha \beta} \\
	10 & L^\alpha L^\beta L^\rho {e}^c Q^\sigma d^c \epsilon_{\alpha \beta}\epsilon_{\rho \sigma} \\
	11a & L^\alpha L^\beta Q^\rho {d}^c Q^\sigma d^c \epsilon_{\alpha \beta}\epsilon_{\rho \sigma} \\
	11b & L^\alpha L^\beta Q^\rho {d}^c Q^\sigma d^c \epsilon_{\alpha \rho}\epsilon_{\beta \sigma} \\
	12a & L^\alpha L^\beta\bar{Q}_\alpha\bar{u}^c \bar{Q}_\beta\bar{u}^c \\
	12b^* & L^\alpha L^\beta\bar{Q}^\rho\bar{u}^c \bar{Q}^\sigma\bar{u}^c\epsilon_{\alpha\beta}\epsilon_{\rho\sigma} \\
	13 & L^\alpha L^\beta\bar{Q}_\alpha\bar{u}^c L^\rho\bar{e}^c\epsilon_{\beta\rho} \\
	14a & L^\alpha L^\beta\bar{Q}_\rho\bar{u}^c Q^\rho d^c\epsilon_{\alpha\beta} \\
	14b & L^\alpha L^\beta\bar{Q}_\alpha\bar{u}^c Q^\rho d^c\epsilon_{\beta\rho} \\
	16 & L^\alpha L^\beta e^c d^c \bar{e}^c \bar{u}^c\epsilon_{\alpha\beta} \\
	19 & L^\alpha Q^\beta d^c d^c \bar{e}^c \bar{u}^c\epsilon_{\alpha\beta} \\
	20 & L^\alpha \bar{Q}_\alpha d^c \bar{u}^c \bar{e}^c \bar{u}^c \\
	66 &  L^\alpha L^\beta H^\rho H^\sigma Q^\eta \bar{H}_\eta d^c \epsilon_{\alpha\rho}\epsilon_{\beta\sigma} \\
	\vspace{1pt} &  \\
	\specialrule{.2em}{.0em}{.3em}
\end{tabular}
\caption{Compilation of relevant dimension-5, 7 and 9 SM invariant $\Delta L = 2$ LNV operators considered in our analysis. The labelling follows Ref.~\cite{Deppisch:2018}; $\dagger$ indicates that $\mathcal{O}_{4_b}$ is Fierz related to $\mathcal{O}_{4_a}$.}
\label{tab:operatorlist}
\end{table}

\subsection{Phenomenology of LNV operators}\label{LNVpheno}
The phenomenology of different LNV operators has been studied with respect to various observables, e.g. the generation of neutrino masses, collider signals or $0\nu\beta\beta$ decay~\cite{Deppisch:2013jxa, Deppisch:2015yqa, Deppisch:2018, deGouvea:2019}. Neutrinoless double beta decay is the primary probe for LNV via light Majorana neutrinos or exotic NP \cite{Deppisch:2012nb}, however, limited to the first generation quarks and charged leptons. At the quark level, the process proceeds by transforming two down-type quarks into two up-type quarks and two electrons.\footnote{We here consider the $0\nu\beta^-\beta^-$ mode with two electrons where searches are most sensitive.} In addition to the standard mass mechanism via light Majorana neutrinos, the decay can thus be mediated by an effective dimension-9 operator with the low energy signature $\bar d \bar d u u e^- e^-$, in a so called short-range contribution \cite{Cirigliano:2018yza, Graf:2018ozy, Deppisch:2020ztt}. Another option is a low energy dimension-6 LNV operator of the form $\bar d u e^- \nu_L$ in combination with a standard Fermi interaction and the exchange of a light neutrino. This is called a long-range contribution and in the SMEFT language it is mediated by dimension-7 and above operators. Current double beta decay searches set a lower limit on the $0\nu\beta\beta$ decay half-life of the order $T_{1/2}^{0\nu\beta\beta} \gtrsim 10^{26}$~yr in various isotopes \cite{Kim:2020vjv}. This can be translated into limits on the operator scales of dimension-7 and dimension-9 operators of the order $\Lambda^{(7)} \sim 10\-- 100$~TeV and $\Lambda^{(9)} \sim 1\-- 10$~TeV \cite{Deppisch:2012nb, Cirigliano:2017djv, Cirigliano:2018yza, Graf:2018ozy, Deppisch:2020ztt}. The most optimistic estimates for the sensitivity of future double beta decay searches predict an improvement by two orders of magnitude, $T_{1/2}^{0\nu\beta\beta} \approx 10^{28}$~yr \cite{Kim:2020vjv}. By its very nature as a nuclear process occurring at an energy scale of $\approx 100$~MeV, $0\nu\beta\beta$ decay probes the relevant LNV operators with first generation quarks and electrons only.

\begin{figure}[t]
	\centering
	\includegraphics[width=0.49\textwidth]{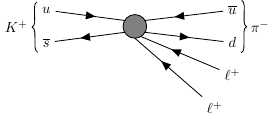}
	\includegraphics[width=0.49\textwidth]{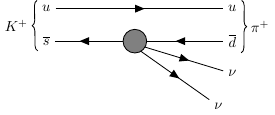}\\
	\hspace{-.72cm}\includegraphics[width=0.43\textwidth]{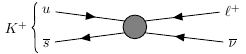}
	\hspace{.88cm}\includegraphics[width=0.43\textwidth]{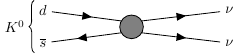}
	\caption{Diagrams of rare lepton number violating kaon decays. Top left: Semi-leptonic decay $K^+ \to \pi^- \ell^+ \ell^+$ with visible lepton number violation in charged leptons. Top right: Semi-leptonic decay $K^+ \to \pi^+ \nu\nu$ involving neutrinos only. Bottom left: Leptonic decay $K^+ \to \ell^+ \bar\nu_\ell$ with one neutrino. Bottom right: Leptonic decay $K^0 \to \nu \nu$ with two neutrinos. }
	\label{fig:kaon-diagrams}
\end{figure}

This limitation does not apply in certain meson systems and LNV operators can be probed in kaon, B-meson, D-meson and $\tau$ decays. Among these, kaon decays will the main focus of this work. At mass dimension-9, it is possible to generate LNV meson decays via a short-range operator $\bar q_d \bar q_d q_u q_u \ell_i^- \ell_j^-$ with two charged leptons involved  \cite{Ali:2001gsa, Liu:2016oph, Quintero:2016iwi}. For instance, such an operator induces the rare decay $K^+ \to \pi^-\ell^+\ell^+$ as shown in Fig.~\ref{fig:kaon-diagrams}~(top left).
This is in direct analogy to $0\nu\beta\beta$ decay, albeit with different quark or lepton generations involved. Likewise, the advantage of these decays is that LNV is directly testable, i.e.\hspace{-1pt} it manifests itself in charged leptons only. However, for a dimension-9 operator, the NP scale $\Lambda$ goes with the fifth power, which leads to a high suppression and current searches for such meson decays can probe operator scales of order $\Lambda^{(9)} \gtrsim (5 - 50)$~GeV only \cite{Quintero:2016iwi}.

At dimension-7, operators will induce fully leptonic LNV decays of the type $K^+ \to \ell^+ \bar\nu$ and $\pi^+ \to \ell^+ \bar\nu$, see Fig.~\ref{fig:kaon-diagrams}~(bottom left). The final state incorporates an anti-neutrino rather than a neutrino and the decay thus violates total lepton number by two units. While the anti-neutrino is not detectable in a given decay, it can be picked up in neutrino oscillation detectors where it produces a positively charged lepton rather than a negatively charged one, as would be expected in a lepton number conserving decay of a meson with positive charge. Neutrino oscillation detectors able to distinguish between the charge of the detected lepton can thus probe such LNV decays \cite{CooperSarkar:1981pb, Bolton:2019wta}.\footnote{The charged lepton promptly produced in the decay is not detected in such a scenario but its charge is inferred from that of the initial meson via an appropriate production mechanism.} We will discuss the limits arising from rare LNV kaon and pion decays in Section~\ref{sec:leptonic-decays}. In Fig.~\ref{fig:kaon-diagrams}~(bottom right), the decay $K^0 \to \nu\nu$ is illustrated. This diagram is included for completeness, but the process is not analyzed further, due to more stringent experimental constraints being put by the semileptonic neutral kaon decays. Invisible decays of the kaon have been studied in \cite{Gninenko:2014sxa}.

Finally, and as our main focus, one can consider the rare semi-leptonic kaon decays $K^+ \to \pi^+ \nu\bar\nu$ and $K_L \to \pi^0\nu \bar\nu$ with neutrinos in the final state. As indicated, the two neutrinos are usually considered to be left-handed only which in the SM corresponds to the processes being lepton number conserving. Instead, we consider the effect of LNV operators on this decay where two neutrinos $\nu\nu$ are being emitted, see Fig.~\ref{fig:kaon-diagrams}~(top right). As the neutrinos are not detectable in rare kaon decay experiments, it is not possible to immediately distinguish between the scenarios of lepton number conservation and violation. The case of LNV thus cannot be excluded in rare kaon decays and hence, it is interesting to consider such a possibility and to study the consequences. For this reason, we will denote the neutrino and anti-neutrino in the same way, and focus on the rare kaon decays $K^+ \to \pi^+\nu\nu$ and $K_L \to\pi^0\nu\nu$.

\begin{figure}
	\centering
	\includegraphics[width=0.32\textwidth]{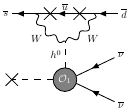}
	\hspace{1.5cm}
	\raisebox{0.07\height}{\includegraphics[width=0.32\textwidth]{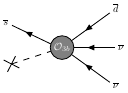}}
	\caption{Left: GIM suppressed long-range contribution of the Weinberg operator $\Op_1$ to the rare kaon decay. Right: Short-range contribution of the operator $\Op_{3b}$.}
	\label{fig:Op3bdiagram}
\end{figure}
In the following, we will assume that the two neutrinos in the final state of the meson decay are of Majorana nature featuring the same lepton number such that the decay violates lepton number by two units. If these possible NP contributions occur at a high energy scale, they can be model-independently described by the $\Delta L = 2$ SMEFT operators introduced above. The lowest LNV operator is the dimension-5 Weinberg operator. However, this operator does not feature a short-range contribution to the rare kaon decay. This can be easily understood, as it does not contain any quark fields that are able to transform the $s$-quark into a $d$-quark. The Weinberg operator, however, can contribute to the rare kaon decay at long-range, see Fig.~\ref{fig:Op3bdiagram}~(left), but, due to the remaining GIM suppression, this diagram is expected to contribute only to a very small extent. The lowest dimension at which a LNV operator can lead to a short-range contribution in rare kaon decays is at mass dimension-7. From Table~\ref{tab:operatorlist}, only a single $\Delta L = 2$ operator $\Op_{3b}$ is able to contribute at dimension-7,
\begin{equation}
\label{eq:7op3b}
    \Op_{3b} = L^\alpha L^ \beta Q^\rho d^cH^\sigma\epsilon_{\alpha \rho}\epsilon_{\beta \sigma}.
\end{equation}
Decomposing it in terms of $SU(2)_L$ component fields, $h_0 d^c s_L \nu_L \nu_L$, one can directly see its contribution to the rare kaon decay as illustrated in Fig.~\ref{fig:Op3bdiagram}~(right).

There is only one way to contract the $SU(2)_L$ indices of operator $\Op_{3b}$ such that it can mediate the rare kaon decay,
\begin{equation}
\label{eq:symbreak1}
\Op_{3b} = L^\alpha_i L^ \beta_j Q^\rho_a d^c_b H^\sigma\epsilon_{\alpha \rho}\epsilon_{\beta \sigma} \rightarrow h^0d_a^cd_{L_b}\nu_{L_i}\nu_{L_j}.
\end{equation}
We have suppressed the spinor indices of the two-component spinor fields on the RHS of Eq.~\eqref{eq:symbreak1} that lead to two possible contractions (indicated by brackets) with Wilson coefficients $c_1$ and $c_2$,
\begin{equation}
\label{eq:postshouten}
h^0d_a^cd_{L_b}\nu_{L_i}\nu_{L_j}  \rightarrow c^{ijab}_1h^0\left(d^{c}_a d_{L_b}\right)\left({\nu_{L_i}}{\nu_{L_j}}\right) + c^{ijab}_2h^0\left({d_a^{c}}{\nu_{L_i}}\right)\left({\nu_{L_j}}{d_{L_b}}\right).
\end{equation}
The two contractions in Eq.~\eqref{eq:postshouten} can be related such that \cite{Dreiner:2008tw}
\begin{align}
\label{eq:postpostshouten}
&c^{ijab}_1h^0\left(d^{c}_a d_{L_b}\right)\left({\nu_{L_i}}{\nu_{L_j}}\right) + c^{ijab}_2h^0\left({d_a^{c}}{\nu_{L_i}}\right)\left({\nu_{L_j}}{d_{L_b}}\right) =\nonumber \\
&\lr{c^{ijab}_1 - \frac{c^{ijab}_2}{2}} h^0\left(d^{c}_a d_{L_b}\right)\left({\nu_{L_i}}{\nu_{L_j}}\right) - \frac{c^{ijab}_2}{2} h^0\left(d^{c}_a\sigma^{\mu\nu} d_{L_b}\right)\left({\nu_{L_j}}\sigma_{\mu\nu}{\nu_{L_i}}\right)\, ,
\end{align}
where $\sigma^{\mu\nu} = \frac{i}{4}(\sigma^\mu\bar\sigma^\nu-\sigma^\nu\bar\sigma^\mu)$, $\sigma^\mu = (\mathbb{1}, \vec{\sigma})$, $\bar\sigma^\mu = (\mathbb{1}, -\vec{\sigma})$, $\vec{\sigma} = (\sigma_x, \sigma_y, \sigma_z)$, with the Pauli matrices $\sigma_i$. We omit the last term in Eq.~\eqref{eq:postpostshouten}, which corresponds to a tensor current contribution that vanishes in case the neutrinos have identical flavour, since this will be the scenario that we focus on. Considering also the hermitian conjugate contribution of $\mathcal{O}_{3b}$ and assuming $c_{1,2} = c_{1,2}^*$, we can rewrite the expression in terms of Dirac spinors, $d = (d_L, \bar{d}^c)^T$, $\nu = (\nu_L, \bar{\nu}_L)^T$,
\begin{align}
\label{eq:postpostshouten2}
&\lr{c^{ijab}_1 - \frac{c^{ijab}_2}{2}} h^0\left[\left(\bar{d}_a d_b\right)\left({\bar{\nu}_i}{\nu_j}\right)+ \left(\bar{d}_a \gamma_5 d_b\right)\left({\bar{\nu}}_i \gamma_5{\nu_j}\right)\right]\,,
\end{align}
where only the scalar interaction will contribute to the decay due to the pseudoscalar nature of the kaon and pion.
In the following, we absorb the Wilson coefficients into the scale of NP $\Lambda_{ijab}$,
\begin{equation}
\label{eq:scaleRedef}
c^{ijab}_1 - \frac{c^{ijab}_2}{2} \rightarrow \frac{1}{\Lambda_{ijab}^3}\,.
\end{equation}
\begin{figure}
	\centering

	\raisebox{0.2\height}{\includegraphics[width=0.32\textwidth]{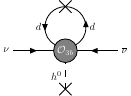}}
	\raisebox{0.2\height}{\includegraphics[width=0.32\textwidth]{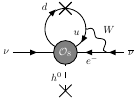}}
	\includegraphics[width=0.32\textwidth]{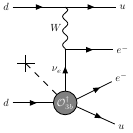}
	\caption{Radiative neutrino mass diagrams for operators $\Op_{3b}$ (left) and $\Op_8$ (center), as well as the contribution to $0\nu\beta\beta$ decay induced by $\Op_{3b}$ (right). For the center diagram, additional mass insertions for the up-type quark and charged lepton are implied.}
	\label{fig:NumassDiagram}
\end{figure}
In general, every higher dimensional operator can contribute to lower dimensional operators through radiative (loop) and symmetry breaking effects, see e.g. Ref.~\cite{deGouvea:2008, Deppisch:2018} in the context of the LNV operators under consideration. Thus each higher dimensional operator that is considered to directly contribute to a specific observable as e.g. rare kaon decay will at the same time contribute to the dimension-5 Weinberg operator. Thus any $\Delta L = 2$ SMEFT operator contributes to a radiatively generated neutrino Majorana mass \cite{Cepedello:2018, Cai:2017}. For example, the operators $\Op_{3b}$ and $\Op_{8}$ generate radiative neutrino masses if the two quark legs are connected via a loop with an additional mass insertion, Higgs or vector boson loop, respectively, see Fig.~\ref{fig:NumassDiagram}. The radiatively generated neutrino mass for $\Op_{3b}$ can be estimated as
\begin{equation}
\label{eq:numassOp3b}
	\delta m_\nu^{(3b)} \approx \frac{y_d}{16\pi^2}\frac{v^2}{\Lambda_{3b}},
\end{equation}
where the appearance of a loop factor and down-type quark Yukawa coupling $y_d$ are evident from Fig.~\ref{fig:NumassDiagram}~(left). Assuming that this contribution saturates the limit on the absolute neutrino mass $m_\nu \lesssim 0.1 - 1$~eV from Tritium decay, $0\nu\beta\beta$ decay and cosmological considerations therefore put a stringent lower limit on the NP scale of the order $\Lambda_{3b} \gtrsim 5\times 10^6$~GeV for the first generation down-type quark Yukawa coupling $y_d = m_d / v$. Similarly, $\Op_{8}$ generates radiatively a contribution to the neutrino mass. While the contributions of the charged Higgs loops cancel each other, the contribution from the vector boson is the dominant one. Such a limit arising from light neutrino masses should be considered indirect in the sense that it assumes that the radiative contribution itself dominates and saturates the bound. Instead, other contributions, at tree level or otherwise, are expected to exist and can destructively compensate each other. As we will see in Sec.~\ref{sec:leptoquarks}, considerations of the flavour structure in UV scenarios can also suppress the radiative neutrino mass contribution.

A similar complementarity can be drawn with constraints from $0\nu\beta\beta$ decay searches. For each operator of rare meson decays in our analysis, we can compare its potential contribution to $0\nu\beta\beta$ decay. However, in this case only a specific flavour combination can be tested, namely the electron contribution. This has been studied in detail in~\cite{Deppisch:2015yqa, Deppisch:2018}. For example, operator $\Op_{3b}$ induces a contribution as shown in Fig.~\ref{fig:NumassDiagram}~(right) that is constrained by the upper limit on the $0\nu\beta\beta$ decay half life to $\Lambda_{3b} \approx 3 \times 10^{5}~\mathrm{GeV}$.

Using the same procedure, higher dimensional LNV operators, other than $\Op_{3b}$ which directly contributes to the rare kaon decay $K\to\pi\nu\nu$, can still be relevant through radiative or symmetry breaking effects. For example, the dimension-9 operator $\Op_{1^{y_d}}$ in Table~\ref{tab:operatorlist}, representing a combination of the singlet Weinberg and down-quark Yukawa operators, yields the effective dimension-6 contribution
\begin{align}
	\frac{1}{\Lambda^2_{K\to\pi\nu\nu}} = \frac{v^3}{\Lambda^5_{1^{y_d}}},
\end{align}
after the Higgs fields acquire their VEV. As another example, the dimension-9 operator $\Op_{11b}$ leads to the contribution
\begin{align}
	\frac{1}{\Lambda^2_{K\to\pi\nu\nu}} = \frac{1}{16\pi^2}\frac{y_d v}{\Lambda^3_{11b}},
\end{align}
with a loop formed by a pair of down-type quark fields that includes a mass insertion with an associated Higgs VEV. At $\Lambda^{(9)} \gtrsim (5 - 50)$~GeV, the scales coming from dimension-9 operators are generally lower than those coming from dimension-7 operators.

\subsection{Implications for baryogenesis mechanisms}
\label{leptogenesis}

Lepton number violating signals cannot only hint towards a possible Majorana nature of neutrinos but they can also help in proceeding in the question of what mechanism generated the baryon asymmetry of our Universe (BAU) quantified in the baryon-to-photon ratio $\eta_B^\text{obs} = (6.20 \pm 0.15) \times 10^{-10}$ \cite{Aghanim:2018eyx}.

While theoretically it is established that the three Sakharov conditions~\cite{Sakharov:1967dj} including $B-L$ violation, $CP$ violation and an out-of-equilibrium mechanism, have to be fulfilled, the underlying mechanism is not yet confirmed. One of the most popular solutions is baryogenesis via leptogenesis~\cite{Fukugita:1986hr}. In this approach a lepton asymmetry is generated in the early Universe via $CP$-violating decays of right-handed neutrinos, which is translated via SM sphaleron processes into the observable baryon asymmetry. However, in order to generate a final lepton asymmetry, so-called washout processes must not be too strong. Otherwise a pre-existing lepton asymmetry could be diminished, leading to an insufficient observable baryon asymmetry. The $\Delta L=2$ LNV processes that we focus on, fall into the category of such washout processes. For a more detailed review on this topic, we refer to \cite{Chun:2017spz, Deppisch:2015yqa, Deppisch:2018}. In case of observing a rare kaon decay pointing towards NP, we can estimate its characteristic energy scale $\Lambda$, which we will discuss in more detail in the next section. Under the assumption that this process violates lepton number, we can derive the scale at which the washout stops being effective. This gives us an indication what this would imply for possible leptogenesis and baryogenesis scenarios. Hereby, we follow the approach as described in Ref.~\cite{Deppisch:2018}. The evolution of the lepton number density can be described by the Boltzmann equation \cite{Buchmuller:2005}
\begin{equation}
\label{eq:Boltzmann}
    H T n_{\gamma}\frac{d\eta_{L}}{dT} = -\lr{\frac{n_{L}n_{i}\dots}{n^{\text{eq}}_{L}n^{\text{eq}}_{i}\dots} - \frac{n_{j}n_{k}\dots}{n^{\text{eq}}_{{j}}n^{\text{eq}}_{{k}}\dots}}\gamma^{\text{eq}}\lr{Li\dots \rightarrow  {j}{k}\dots} + \text{permutations.}
\end{equation}
Here, $n_i$ is the number density of particle $i$, with $\eta_i \equiv n_i/n_\gamma$, where $n_\gamma$ is the number density of photons. Furthermore, $H$ is the Hubble constant. A superscript \textit{eq} indicates an equilibrium distribution, and $\gamma^{\text{eq}}$ denotes the equilibrium reaction density. The dots represent any other particles that appear in the relevant process. In the following we will consider all processes that one single operator at a time can generate. Hence, \textit{permutations} indicates all other possible orderings of fields that an operator (including the Hermitian conjugate) can generate.
Estimating the equilibrium reaction density $\gamma^\text{eq}$ according to \cite{Deppisch:2018} and expressing the number densities using chemical potentials and relating those of SM fields through equilibration of the Yukawa couplings and sphaleron processes \cite{Harvey:1990}, the relevant Boltzmann equation is given with respect to the total asymmetry of the lepton doublet number density,
\begin{equation}
    \label{eq:secondBoltzmann}
    \frac{d\eta_{\Delta L}}{dz} = - \frac{\eta_{\Delta L}}{z}c'_D\frac{\Lambda_{\text{Pl}}}{\Lambda}\lr{\frac{T}{\Lambda}}^{2D-9},
\end{equation}
with $\eta_{\Delta L} \equiv \eta_L - \eta_{\bar L}$. The dimensionless coefficient $c_D'$ is determined according to \cite{Deppisch:2018} by relating the different chemical potentials and considering all possible permutations of the process as indicated in Eq.~\eqref{eq:Boltzmann} \cite{Deppisch:2018}. The Planck scale is defined as $\Lambda_\text{Pl} = \sqrt{8\pi^3 g_*/90}T^2/H$ assuming a flat Friedmann-Robertson-Walker universe, where $g_*$ is the number of relativistic degrees of freedom in the SM. Solving the Boltzmann equation in Eq.~\eqref{eq:secondBoltzmann} shows that even a large asymmetry $\eta_{\Delta L} = 1$ at an initial temperature $T$ is washed out to a level below the observed baryon asymmetry $\eta^\text{obs}_B$ in the temperature range
\begin{align}
\label{eq:O1scale}
	\hat\lambda \lesssim T \lesssim \Lambda.
\end{align}
Here, the upper limit is imposed by the effective operator description breaking down at and above $\Lambda$. The lower limit is given by \cite{Deppisch:2018}
\begin{align}
\label{eq:O11bscale}
	\hat\lambda = \left[(2D-9)\ln\left(\frac{10^{-2}}{\eta_B^\text{obs}}\right)\lambda^{2D-9}
	+v^{2D-9}\right]^{1/(2D-9)},
\end{align}
with $\lambda = \Lambda(\Lambda/(c'_D \Lambda_\text{Pl}))^{1/(2D-9)}$. Effectively this means that an asymmetry $\eta_{\Delta L} = 1$ generated at an energy scale above $\hat\lambda$ in some mechanism will be erased by the presence of the given dimension-$D$ LNV operator. Applying this to our operators in question, this washout scale is shown in Table~\ref{tab:d7mesondecay2}, and it is very close to the electroweak symmetry breaking scale for all listed LNV observables at their current experimental limits. Taking the NP scale as the upper limit for the washout for each operator, we have a range over which the washout is effective. The observation that the washout is effective over a wide energy range for all dimension-7 operators in the meson decays tells us that any NP in these rare decays would reveal a new contribution to the washout in a leptogenesis scenario, assuming the neutrino has a Majorana mass.

%% file: rarekaon.tex
\section{Rare Kaon Decays}\label{sec:rarekaon}

In the following, we discuss in detail the possibility of LNV interactions contributing to rare kaon decays. For a direct comparison, we first review the expected SM contribution, in order to focus then on a possible NP contribution due to an underlying LNV operator. Such contributions were discussed in \cite{Li:2019fhz}, but we specifically focus on how the different currents associated with a LNV or LNC interaction lead to kinematic differences. This has an impact on the sensitivity of past and ongoing experiments and opens the door to disentangle LNC and LNV contributions in the future. We derive limits on the NP scale for different LNV operators and complement our analysis by considering fully leptonic meson decays and other complementary probes for LNV.

\subsection{$K\to\pi\nu\bar{\nu}$ within the Standard Model}
\label{RareKaonSM}
\begin{figure}
	\centering
	\includegraphics[width=0.47\textwidth]{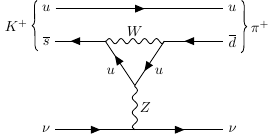}
	\includegraphics[width=0.47\textwidth]{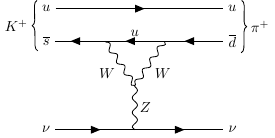}\\[4mm]
	\includegraphics[width=0.47\textwidth]{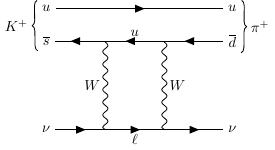}
	\caption{Contributions to $K^+\to\pi^+\nu\bar{\nu}$ within the SM. A summation over the internal quark and lepton flavours is implied.}
	\label{fig:SMdiagrams}
\end{figure}

We focus on the two so-called golden modes $K^+\to\pi^+\nu\bar{\nu}$ and $K_L\to \pi^0\nu\bar{\nu}$. The former proceeds in the SM via the electroweak penguin and box diagrams as depicted in Fig.~\ref{fig:SMdiagrams} \cite{Inami:1981}. Due to loop and GIM suppression~\cite{Glashow:1970gm}, the branching ratio of this decay mode is very small.

When replacing the spectator $u$-quark in Fig.~\ref{fig:SMdiagrams} by a $d$-quark, the neutral decay $K_0\to \pi^0\nu\bar{\nu}$ is described. With $K^0$ and $\bar{K}^0$ transforming into each other under CP conjugation, only their linear combination leads to $CP$ eigenstates. Their mass eigenstates $K_L$ and $K_S$ contain a small admixture $\epsilon$ of the state with opposite CP parity~\cite{DAmbrosio:1996lam},
\begin{equation}
\label{eq:klong}
	|K_L\rangle =
	\frac{1}{\sqrt{2 + 2|\epsilon|^2}}
	\left[(1 + \epsilon)|K^0\rangle + (1 - \epsilon)|\bar{K}^0\rangle\right]\,,
\end{equation}
where $\epsilon$ is a small experimentally determined parameter of $\mathcal{O}(10^{-3})$ quantifying the indirect CP violation in the mixing.

The theoretically predicted branching ratios of the above SM rare decays can be parametrized as \cite{Buchalla:1998ba,He:2018},
\begin{align}
\label{eq:SMBRa}
	\text{BR}(K^+\to\pi^+\nu\bar\nu) &=
	\tilde\kappa^+
	\left[\left(\frac{\Im(V_{ts}^* V_{td}X_t)}{\lambda^5}\right)^2
	    + \left(\frac{\Re(V_{cs}^* V_{cd})}{\lambda}P_c
	          + \frac{\Re(V_{ts}^* V_{td}X_t)}{\lambda^5}\right)^2\right], \\
\label{eq:SMBRb}
	\text{BR}(K_L\to\pi^0\nu\bar\nu) &=
	\kappa_L \left(\frac{\Im(V_{ts}^* V_{td}X_t)}{\lambda^5}\right)^2.
\end{align}
Here, $X_t = 1.48$ and $P_c = 0.404$ are dimensionless quantities corresponding to higher order effects of top and charm quarks, respectively, $V_{ij}$ are the usual CKM matrix elements and $\lambda\approx 0.225$ is the Wolfenstein parameter. The quantities $\tilde\kappa^+ = 0.517\times 10^{-10}$ and $\kappa_L = 2.23\times 10^{-10}$ include the hadronic matrix elements determined with a small uncertainty due to their relation to the well measured branching ratios of the more rapid decays $K^+\to\pi^0 e^+\nu$ and $K_L\to\pi^-e^+\nu$, respectively~\cite{Bryman:2005xp, Mescia:2007}. This allows to theoretically predict the branching ratios of the SM rare kaon decay to a high precision \cite{Buras:2015, PDG:2018},
\begin{align}
\label{eq:SMBRTHvaluea}
	\text{BR}(K^+\to\pi^+\nu\bar{\nu})_\text{SM} &= \left(8.4 \pm 1.0\right)\times 10^{-11}, \\
\label{eq:SMBRTHvalueb}
	\text{BR}(K_L\to\pi^0\nu\bar{\nu})_\text{SM} &= \left(3.4 \pm 0.6\right)\times 10^{-11}.
\end{align}
The uncertainty of the predicted SM value is mainly limited by the experimental accuracy of the CKM matrix elements~\cite{Buras:2015}. Due to this theoretical cleanliness~\cite{Buras:2008}, the rare kaon decays will provide us with an excellent probe for NP.

It is important to stress that the decay branching ratio of $K_L\to\pi^0\nu\bar\nu$ in Eq.~\eqref{eq:SMBRb} is proportional to the imaginary part of the CKM matrix elements, hence requiring CP violation in the SM, with a sub-dominant CP conserving contribution occurring only at second order \cite{Buchalla:1998}. This as a consequence of the two final state neutrinos forming a CP-odd state if arising from a left-handed vector current unless there is neutrino flavour violation \cite{Grossman:1997,Grossman:2003rw}. 

As evident from Eqs.~\eqref{eq:SMBRTHvaluea} and \eqref{eq:SMBRTHvalueb}, the rare decay branching ratio of $K_L$ is smaller than that of $K^+$. Even under the presence of exotic contributions, the branching ratios are related as
\begin{equation}
\label{eq:grossman-nir}
	\text{BR}(K_L\to \pi^0\nu\bar\nu) < 4.4\times\text{BR}(K^+\to \pi^+\nu\bar\nu).
\end{equation}
This is called the Grossman-Nir (GN) bound~\cite{Grossman:1997}, and it applies to a wide range of NP contributions. Breaking of the GN bound is possible if operators with exotic isospin $\Delta I = 3/2$~\cite{He:2020jzn, He:2020jly} are considered, hence requiring exotic dark particles \cite{Fuyuto:2014cya, Fabbrichesi:2019bmo, Dev:2019hho, Hostert:2020gou, 1796725}. The bound may also be effectively modified if other decay modes are considered, e.g. $K\to\pi X$ with an exotic scalar.

\subsection{Experimental searches}
\label{RareKaonExperiment}

Currently, the most stringent and confirmed bounds on the branching ratios of the rare kaon decays are given by the E949 experiment for $K^+$ \cite{Artamonov:2009} and by the KOTO experiment for $K_L$ \cite{Ahn:2019, Kitahara:2019}.

\paragraph{The~E949~experiment} has searched for the decay $K^+ \to \pi^+ +$~nothing at the Brookhaven National Laboratory, using stopped kaons \cite{Artamonov:2009}. Interpreting the result in terms of the SM contribution, the E949 experiment arrives at
\begin{align}
\label{eq:SMBREXPvalueE949}
    \text{BR}(K^+\to\pi^+\nu\bar\nu)_\text{E949, vector}
    = \left(1.73^{+1.15}_{-1.05}\right)\times 10^{-10},
\end{align}
with an upper limit
\begin{align}
\label{eq:SMBREXPvalueE949}
    \text{BR}(K^+\to\pi^+\nu\bar\nu)_\text{E949, vector}
    < 3.35\times 10^{-10},\quad\text{at 90\% CL},
\end{align}
in agreement with the SM prediction in Eq.~\eqref{eq:SMBRTHvaluea}.
The E949 experiment also reports an upper limit on the branching ratio in case of a scalar current, given by
\begin{align}
\label{eq:E949scalarBR}
    \text{BR}(K^+\to\pi^+\nu\bar\nu)_\text{E949, scalar} < 21\times 10^{-10},
    \quad\text{at 90\% CL}.
\end{align}
The E949 experiment uses two signal regions $\pi\nu\overline{\nu}(1)$ and $\pi\nu\overline{\nu}(2)$, which correspond to kinematic cuts in the measured momentum range and kinetic energy of the pion.
In the following, we will, for simplicity, approximate
the signal regions by performing cuts in the momentum $p_\pi$ only such that $211 \text{ MeV} < p_\pi < 229 \text{ MeV}$ ($\pi\nu\overline{\nu}(1)$) and $140 \text{ MeV} < p_\pi < 199 \text{ MeV}$ ($\pi\nu\overline{\nu}(2)$).

\paragraph{The NA62 experiment} is an ongoing effort searching for the decay-in-flight $K^+ \to \pi^+ +$~nothing at CERN \cite{CortinaGil:2018fkc,CortinaGil:2020vlo}. Both the final pion momentum and the missing squared-energy $s$ are observables. For the NA62 experiment, the signal regions 1 and 2 correspond to $0 < s < 0.01~\text{GeV}^2$ and $0.025~\text{GeV}^2 < s < 0.068~\text{GeV}^2$, respectively, with both signal regions also being constrained in the pion momentum by $15~\text{GeV} < p_\pi < 35~\text{GeV}$. NA62 is expected to probe the decay at SM sensitivity in the near future \cite{Martellotti:2018}. The most stringent upper limit put by NA62 is\footnote{A central value of $\text{BR}(K^+\to\pi^+\nu\bar\nu)_\text{NA62} = \lr{11.0^{+4.0}_{-3.5}\pm 0.3}\times 10^{-11}$ ($3.5\sigma$ significance) \cite{NA62talkICHEP} based on 17 observed events with an estimated background of 5.3 events has recently been reported by the NA62 collaboration.}
\begin{align}
\label{eq:SMBREXPvalueNA62}
	\text{BR}(K^+\to\pi^+\nu\bar\nu)_\text{NA62} < 1.78\times 10^{-10},
	\quad\text{at 90 \% CL}.
\end{align}
This is in agreement with both the SM prediction and the E949 result. In the future, the NA62 experiment aims to probe the rare kaon decay with 10\% precision \cite{Kleimenova:2019pcu}. Based on this prescription, we estimate a projected upper limit
\begin{align}
\label{eq:SMBREXPvalueNA62future}
	\text{BR}(K^+\to\pi^+\nu\bar\nu)_\text{NA62}^{\text{future}} \lesssim 1.11\times 10^{-10},
	\quad\text{at 90\% CL}.
\end{align}

\paragraph{The~KOTO~experiment} is searching for the decay  $K_L \to \pi^0 +$~nothing at J-PARC \cite{Ahn:2019}. Like NA62, KOTO uses decay-in-flight techniques, but their observables differ due to the different particles involved. The decay of the final pion into two photons occurs within the kaon beam, and the position of the decay along the beam, as well as the transverse momentum of the pion, are reconstructed from measurements of the two photons \cite{Ahn:2019}. The KOTO experiment has a single signal region which is defined by cuts in the transverse momentum $p_T$ of the final state pion, and the location of the decay $Z_\text{vtx}$ along the beam axis, where the lower cut in transverse momentum depends on $Z_\text{vtx}$. The kaon momentum distribution peaks at $1.4$ GeV, with a broad spectrum ranging from 0 GeV to approximately 5 GeV \cite{Masuda:2015eta}. The branching ratio obtained by KOTO is
\begin{align}
\label{eq:2015valueKOTO}
	\text{BR}(K_L\to\pi^0\nu\bar\nu)_\text{KOTO} < 3.0\times 10^{-9},
	\quad\text{at 90\% CL}.
\end{align}
Interpreting the result using the 2016-2018 data in terms of the SM contribution, KOTO finds
\begin{align}
\label{eq:SMBREXPvalueKOTO}
	\text{BR}(K_L\to\pi^0\nu\bar\nu)_\text{KOTO} = \left(2.1^{+4.1}_{-1.7}\right)\times 10^{-9},
	\quad\text{at 95\% CL}.
\end{align}
As evident from Eq.~\eqref{eq:SMBREXPvalueKOTO}, the KOTO result is barely in agreement with the GN bound arising from Eq.~\eqref{eq:SMBREXPvalueE949}, with the KOTO central value higher and outside the GN bound. This makes the KOTO result difficult to interpret in terms of NP. Furthermore, the 2016-2018 data of KOTO is only preliminary and subject to further analysis \cite{KOTOtalk}.

We emphasize that the above quoted limits and measurements, except for the limit in Eq.~\eqref{eq:E949scalarBR}, were derived assuming the SM contribution or more generally an effective operator with the same Lorentz structure as the SM contribution. Due to the kinematic selection criteria and acceptances, all experiments only probe a part of the available phase space and the theoretical decay rate and thus branching ratio can only be determined assuming a given differential phase space distribution. To obtain an experimental branching ratio for different currents, it is not enough to only consider the phase space cuts, due to the difference in expected background in the parts of the phase space. A statistical analysis of the signal events is therefore needed, which is beyond the scope of this work. As we will see below, LNV contributions involving scalar currents have a considerably different distribution and the derived bounds on the branching ratio will be modified.

\subsection{$K\to\pi\nu\nu$ beyond the Standard Model}
\label{kaontheory}

In order to study the impact of potential new, LNV physics on the rare kaon decay $K\to\pi\nu\nu$, we parametrize its matrix element in terms of the effective low energy interaction arising from the operator $\Op_{3b}$ in Eq.~\eqref{eq:7op3b},
\begin{align}
\label{eq:matrixelementfirst}
	\iM =
	\frac{v}{\Lambda^3_{ijsd}}
	\bra{\pi \nu_i \nu_j} \bar ds \overline{\nu}_i \nu_j\ket{K}.
\end{align}
Here, the fermion 4-component fields are defined as $d = (d_L, \bar{d}^c)^T$, $\nu = (\nu_L, \bar{\nu}_L)^T$. The quark flavour indices are fixed to $s$ and $d$ by the kaon and pion quark flavour content.\footnote{Note that in Eq.~\eqref{eq:matrixelementfirst}, the initial $K$ denotes a general kaon, and the quark flavours $\bar ds$ in the operator should be adjusted accordingly for a $K^+/K^0$ or $K^-/\bar K^0$ initial state.} The indices $i$, $j$ label the different neutrino states. In calculating the amplitudes of the rare kaon decays, a form-factor approach is used in this section. It is possible to also perform these calculations using chiral perturbation theory \cite{Li:2019fhz}.

Given the field ordering in Eq.~\eqref{eq:postpostshouten}, we can replace the meson component of the matrix element in Eq.~\eqref{eq:matrixelementfirst} with the corresponding scalar form factor,
\begin{align}
\label{eq:nuclearME}
	\iM = \frac{v}{\Lambda^3_{ijsd}}
	\bra{\pi(p')}\bar ds\ket{K(p)}\bar\nu_i(k) \nu_j(k').
\end{align}
This yields the squared matrix element

\begin{align}
	|\mathcal{M}|^2 = \frac{v^2}{\Lambda^6_{ij sd}}
	\left(\frac{m_K^2 - m_\pi^2}{m_s - m_d}f^K_0(s)\right)^2 s.
\end{align}
The squared meson momentum transfer and, equivalently, the invariant mass-squared of the neutrino system is $s = (p - p')^2 = (k + k')^2$. The scalar form factor $f_0^K(s)$ is defined through~\cite{Shi:2019, Colangelo:2019}
\begin{align}
\label{eq:scalarff}
	\bra{\pi(p')}\bar{d} s\ket{K(p)} = \frac{m_K^2 - m_\pi^2}{m_s - m_d}f^K_0(s),
\end{align}
with the $s$, $d$ quark masses $m_s = 95$~MeV and $m_d = 4.7$~MeV, and the form factor itself is given by \cite{Mescia:2007, Colangelo:2019}
\begin{align}
\label{eq:kaonformfactors}
	f^K_0(s) = f^K_+(0)\left(1 + \lambda_0\frac{s}{m_\pi^2}\right),
\end{align}
with $\lambda_0 = 13.38\times 10^{-3}$ and the factors at zero momentum transfer,
\begin{align}
\label{eq:fK0}
	f^{K^+}_+(0) = 0.9778, \quad f^{K_L}_+(0) = 0.9544,
\end{align}
for the decay of $K^+$ and $K_L$, respectively. The scalar form factor can be derived from the vector form factor~\cite{Shi:2019, Colangelo:2019} using the equations of motion for the quarks \cite{Sakaki:2013}. Also note that the pseudo-scalar part in Eq.~\eqref{eq:nuclearME} vanishes, $\bra{\pi}\bar{d}\gamma_5 s\ket{K} = 0$, as the transition of the pseudo-scalar kaon to a pseudo-scalar pion is parity conserving\footnote{In our analysis we consider only pseudoscalar pion final states. In principle, a similar study could also be conducted for vector mesons in the final state. In this case, only the pseudoscalar current would contribute while the scalar current would vanish. However, the lightest vector meson, $\rho$, is too massive to be produced in kaon decays. It can be produced e.g. in the rare decay $B \to \rho\nu\nu$ and puts a bound on the NP scale in $B$-meson decays ($\text{BR}(B^+ \rightarrow \rho^+\nu\nu) < 3.0 \times 10^{-5}$ at 95\% CL \cite{PDG:2018}). However, the bound from $B \rightarrow \pi\nu\nu$ is more stringent ($\text{BR}(B^+ \rightarrow \pi^+\nu\nu) < 1.4 \times 10^{-5}$ at 95\% CL \cite{PDG:2018}). As this is the case for most pseudoscalar meson decays into vector meson final states, we do not consider them in our analysis.}.

Given the matrix element Eq.~\eqref{eq:nuclearME}, the differential decay rate may be expressed in terms of the invariants $s$ and $t = (k' + p')^2$ as
\begin{align}
\label{eq:decayrate}
	\frac{\Gamma\left(K\to\pi\nu_i\nu_j\right)}{ds\,dt} &=
	\frac{1}{1+\delta_{ij}}\frac{1}{(2\pi)^3}\frac{1}{32m_K^3} |\overline{\mathcal{M}}|^2,
	\nonumber\\
	&= \frac{1}{1+\delta_{ij}}\frac{1}{(2\pi)^3}\frac{1}{32m_K^3}
	   \frac{v^2}{\Lambda_{ijsd}^6}\left(\frac{m_K^2 - m_\pi^2}{m_s - m_d}\right)^2|f^K_0(s)|^2s,
\end{align}
where the factor $1/(1+\delta_{ij})$ is included to account for two identical neutrinos. The phase space is described by the intervals $t \in [t^-, t^+]$ and $s \in [0,(m_K - m_\pi)^2]$ with
\begin{gather}
	t^{\pm} = m_\pi^2 - \frac{1}{2}\left(s-\left(m_K^2 - m_\pi^2\right)
	\mp \sqrt{\lambda\left(s,m_K^2,m_\pi^2\right)}\right),
\end{gather}
where $\lambda\lr{a,b,c}$ is the K\"{a}ll\'{e}n function.

Performing the integrals and using the total $\Gamma_{K+}$ decay width $\Gamma_{K^+} = 5.32\times 10^{-17}$~GeV, the $K^+\to\pi^+\nu\nu$ branching ratio may be expressed as
\begin{align}
	\text{BR}_\text{LNV}(K^+\to\pi^+\nu_i\nu_j) = 10^{-10} \left(\frac{19.2~\text{TeV}}{\Lambda_{ijsd}}\right)^6.
\end{align}
Likewise, using the total $\Gamma_{K_L}$ decay width $\Gamma_{K_L} = 1.29\times 10^{-17}$~GeV, the $K_L\to\pi^0\nu\nu$ branching ratio is expressed as
\begin{align}
	\text{BR}_\text{LNV}(K_L\to\pi^0\nu_i\nu_j) = 10^{-10}\left(\frac{24.9~\text{TeV}}{\Lambda_{ijsd}}\right)^6,
\end{align}
where we assume the scale $\Lambda_{ijsd}$ to be real. In the decay of $K_L$ via a scalar current, apart from the contribution from the small parameter $\epsilon$ in Eq.~\eqref{eq:klong}, the amplitude is proportional to the real part of the coupling, as opposed to the imaginary part in the case of a vector current. This can be seen from the transformation properties between a $\ket{K^0}$ and $\ket{\bar K^0}$ state for the different currents,
\begin{equation}
\label{eq:trafo}
\begin{aligned}
	\bra{\pi^0}\bar{d}\pl s\ket{\bar K^0} &= \phantom{-}\bra{\pi^0}\bar{s}\pl d\ket{K^0},\\
	\bra{\pi^0}\bar{d}\gamma^\mu\pl s\ket{\bar K^0} &= -\bra{\pi^0}\bar{s}\gamma^\mu\pl d\ket{K^0}.
	\end{aligned}
\end{equation}
Adding the $\ket{K^0}$ and $\ket{\bar K^0}$ contributions, the amplitude for the $K_L$ decay is \cite{Buras:1998raa}
\begin{equation}
\begin{aligned}
	\iM\left(K_L\to\pi^0\nu\overset{\brobor}{\nu}\right) = \frac{1}{\sqrt{2+2|\epsilon|^2}}\Big(&F(1+\epsilon)\bra{\pi^0}C\ket{K^0}\\
	+&F^*(1-\epsilon)\bra{\pi^0}C\ket{\bar K^0}\Big)\nu C\overset{\brobor}{\nu},
	\end{aligned}
\end{equation}
where the current $C$ is either of the form $V-A$ or $S-P$ as indicated in Eq.~\eqref{eq:trafo}. Hence, the imaginary part of the coefficient $F$ containing the underlying physics parameters is picked out for the SM $V-A$ case, whereas the real part remains for $S-P$, applicable to our LNV mode.

The SM contribution can be derived using the effective dimension-6 interaction
\begin{align}
\label{eq:smop2}
	\mathcal{L}_\text{SM}^{K\to\pi\nu\bar\nu} = \frac{1}{\Lambda^2_\text{SM}}\sum_{i=1}^3
	\left(\bar\nu_i\gamma^\mu\pl\nu_i\right)\left(\bar{d}\gamma_\mu\pl s\right).
\end{align}
Using the same formalism as above, the matrix element for kaon decay is given by
\begin{align}
	\iM = \bra{\pi(p')\nu_i(k)\bar\nu_i(k')}\mathcal{L}_\text{SM}^{K\to\pi\nu\bar\nu}\ket{K(p)},
\end{align}
yielding the squared matrix element
\begin{align}
\label{eq:LNCmelement}
	|\mathcal{M}|^2 = \frac{6}{\Lambda_\text{SM}^4}
	\left[m_K^2\left(t-m_\pi^2\right) - t\left(s + t - m_\pi^2\right)\right]f^K_+(s)^2.
\end{align}
Here, the form factor arising from the quark vector current is given by
\begin{align}
\label{eq:kaonformfactors}
	f^K_+(s) = f^K_+(0)\left(1 + \lambda_+'\frac{s}{m_\pi^2} + \lambda_+''\frac{s^2}{m_\pi^4}\right).
\end{align}
The constants are $\lambda_+' = 24.82\times10^{-3}$, $\lambda_+'' = 1.64\times10^{-3}$ and the zero-momentum transfer values are as in Eq.~\eqref{eq:fK0}. Integrating over the phase space as described above yields the SM branching ratios in Eqs.~\eqref{eq:SMBRa} and \eqref{eq:SMBRb} where the effective operator scales can be matched with the loop calculation result yielding $|\Lambda_\text{SM}^{K+}| = 8.5$~TeV and $|\Im(\Lambda_\text{SM}^{K_L})| = 15.4$~TeV.

The branching ratios of $K^+\to\pi^+\nu\overset{\brobor}{\nu}$ and $K^0\to\pi^0\nu\overset{\brobor}{\nu}$ under the presence of both the LNV and the SM contribution are then given by
\begin{equation}
    \label{eq:Scale2BR}
    \text{BR}(K\to\pi\nu\overset{\brobor}{\nu}) =
    \text{BR}_\text{SM}(K\to\pi\nu\bar\nu) + \sum_{i\leq=j}^3\text{BR}_\text{LNV}(K\to\pi\nu_i\nu_j).
\end{equation}
The interference between the SM and LNV contribution is negligible, being suppressed by the small neutrino mass.

\subsection{Kinematic distributions in $K\to\pi\nu\nu$}
\label{kaonpheno}
As emphasized above, the LNV contribution to the rare kaon decays proceeds via a scalar current in contrast to a vector current in the SM. This modifies the differential decay distributions as observed experimentally. Once the decays have been positively observed this can be potentially used to distinguish between the different contributions or probe an exotic admixture within a dominant SM distribution. The different distributions are also important in setting limits on exotic contributions as the NA62, E949 and KOTO experiments have different kinematic acceptances and can probe only a fiducial subset of the whole phase space. The derived limits on or measurements of the branching ratios are therefore always dependent on the assumed kinematic distribution.

\begin{figure}
	\centering
	\includegraphics[width=0.51\textwidth]{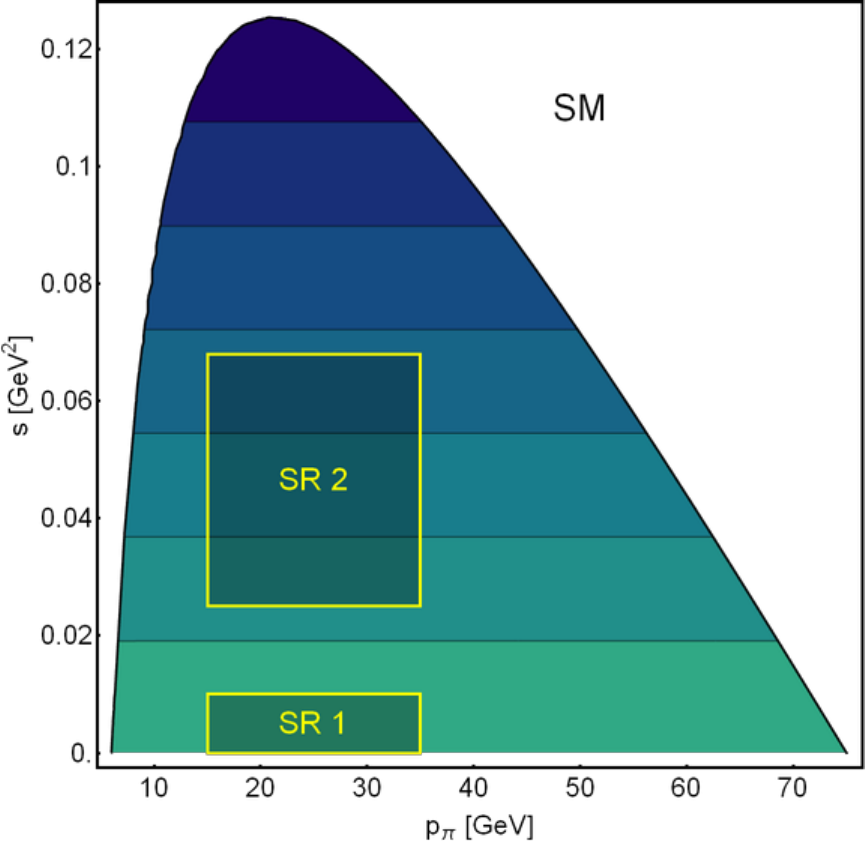}
	\hspace{-10mm}
	\includegraphics[width=0.51\textwidth]{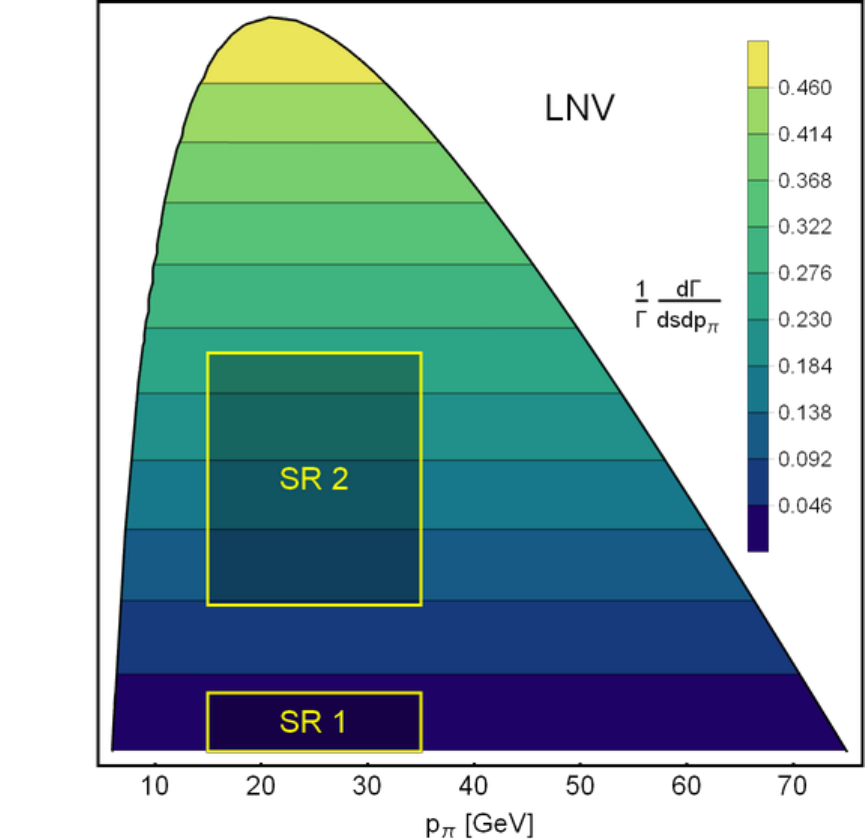}
	\caption{Normalized double differential decay width with respect to the squared missing energy $s$ and pion momentum $p_\pi$ for the SM decay $K^+\to\pi^+\nu\bar\nu$ (left) and the LNV decay $K^+\to\pi^+\nu\nu$ (right). The pion momentum is measured in the lab frame where the kaon has the momentum $|\mathbf{p}_{K^+}| = 75$~GeV and the shaded areas correspond to the two signal regions of the NA62 experiment.}
	\label{fig:contour}
\end{figure}
In Eqs.~\eqref{eq:nuclearME} and \eqref{eq:LNCmelement} we gave the kaon decay matrix elements of the exotic LNV operator and the SM case, respectively, expressed in terms of the invariants $s = (p-p')^2$ and $t = (k'+p')^2$. As described in Eq.~\eqref{eq:decayrate}, this determines the differential decay rate. In Fig.~\ref{fig:contour} we show the double differential $K^+\to\pi^+\nu\bar\nu$ decay width as a function of the pion momentum $p_\pi = |\mathbf{p}_\pi|$ and missing squared-energy $s$ in the lab frame of the NA62 experiment with kaon momentum $|\mathbf{p}_{K^+}| = 75$~GeV, for both the SM (left) and LNV case (right). The upper edges of the phase space correspond to configurations where the direction of the pion momentum is parallel to the kaon beam. We also indicate the two signal regions (SR1, SR2) of the NA62 experiment as defined in Sec.~\ref{RareKaonExperiment}. They are designed to minimize the background, which we will indicate below. In this representation, the kinematic distributions in the SM and LNV case are strikingly different, with the former peaking at $s = 0$ and the latter at the maximum missing squared-energy $s_\text{max} = \lr{m_K-m_\pi}^2$. 
This can be understood from angular momentum considerations with the left-handed vector and scalar currents involved.

\begin{table}
	\centering
	\begin{tabular}{l|cc}
		\specialrule{.2em}{.3em}{.0em}
		\text{Experiment} & \text{SM (vector)} & \text{LNV (scalar)}\\[1mm]
		\hline
		\text{NA62 SR 1} & 6\% & 0.3\% \\[1mm]
		\text{NA62 SR 2} & 17\% & 15\% \\[1mm]
		\hline
		\text{E949 $\pi\nu\overline{\nu}(1)$} & 29\% & 2\% \\[1mm]
		\text{E949 $\pi\nu\overline{\nu}(2)$} & 45\% & 38\% \\[1mm]
		\hline
		\text{KOTO} & 64\% & 30\% \\[1mm]
		\specialrule{.2em}{.0em}{.3em}
	\end{tabular}
	\caption{Percentage of the differential partial width $\Gamma(K\to\pi\nu\nu)$ that falls within the kinematic cuts corresponding to the signal regions of the NA62, E949, and KOTO experiments, as defined in Section~\ref{RareKaonExperiment}, for both vector and scalar currents.}
	\label{tab:ScVePercent}
\end{table}
The signal regions define the fiducial part of the phase space covered by the experiment. In Table~\ref{tab:ScVePercent}, the fraction of the total width that falls within the two NA62 signal is shown. The fractions are obtained by integrating over the differential decay rate within each kinematic signal region. The percentages  indicate how sensitive the experiment is to a specific mode but we stress that it does not directly allow to set limits. For this, a dedicated statistical analysis is needed. From Table~\ref{tab:ScVePercent} it is apparent that NA62 is expected to be more sensitive to a vector current than a scalar current, with a total coverage SR~1 + SR~2 $\approx 23\%$ in the SM compared to SR~1 + SR~2 $\approx 15\%$ in the LNV case. The different sensitivities of the two signal regions SR~1 and SR~2 can also be used to distinguish between vector and scalar currents as the ratio of number of events between the two regions is predicted from kinematics. Specifically, the ratio SR~1 : SR~2 $\approx 0.02$ is very small in the LNV case compared to SR~1 : SR~2 $\approx 0.35$ for the SM.

\begin{figure}
	\centering
	\includegraphics[width=0.65\textwidth]{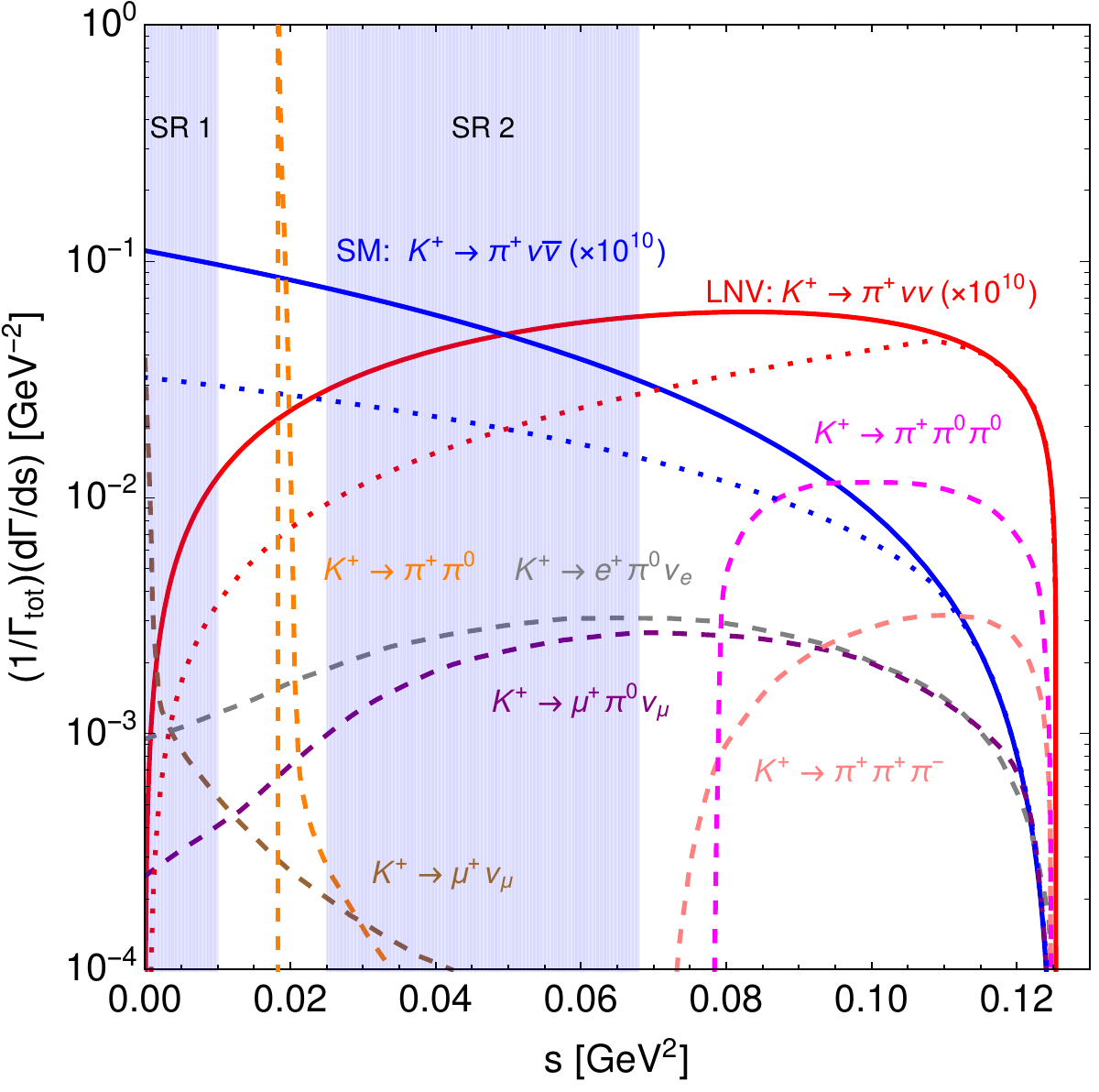}
	\caption{Differential decay width with respect to the squared missing energy $s$. The red and blue solid lines correspond to the LNV and SM decays $K^+\to\pi^+\nu\nu$ and $K^+\to\pi^+\nu\bar\nu$, respectively, where both differential decay widths have been multiplied with a factor $10^{10}$ for visibility. The corresponding red and blue dotted lines show the same but with the additional constraint that only the pion momentum range of the NA62 signal regions has been integrated over.
	The dashed lines show distributions of relevant background processes, and the two light blue areas are the two signal regions at the NA62 experiment. For the background processes, $s$ is defined under the assumption that the final state charged particle is a $\pi^+$.}
	\label{fig:backgrounds}
\end{figure}
In Fig.~\ref{fig:backgrounds} we show the partial decay width in the SM and LNV case as a function of $s$, calculated by integrating over the pion momentum in Fig.~\ref{fig:contour}, in comparison to the relevant background for NA62 as given in \cite{CortinaGil:2018fkc}. Here, $s$ denotes the energy carried away by the neutrino pair, which is not measured directly, but inferred from the pion energy and direction. For the SM mode, we have used the SM scale fixed by Eq.~\eqref{eq:SMBRTHvaluea}, while the LNV operator scale is chosen such that both decays have equal total decay widths. As the SM and LNV rare kaon decay widths are much smaller than those of the background processes, both are multiplied by a factor $10^{10}$ for better visibility. However, within the signal regions further efficient background rejection is applied, as the main background arises from $K^+ \to e^+ \pi^0 \nu_e$ and $K^+ \to \mu^+ \pi^0 \nu_\mu$, where particle identification and photon rejection can be used \cite{CortinaGil:2018fkc}. Photon rejection reduces the accepted signal of background events with $\pi^0$ in the final state, as the $\pi^0$ quickly decays into two high energy photons. Assuming the charged particle in the final state is a $\pi^+$, the kinematics of decays with misidentified $\mu^+$ or $e^+$ final states will appear to violate energy conservation. From subsequent measurements of the charged particle momentum, $\pi^+$ particles are then identified based on their kinematics. In Fig.~\ref{fig:backgrounds}, we also show the signal region intervals in $s$ of the NA62 experiment shaded in blue. In addition, the red and blue dotted curves give the SM and LNV differential decay rates calculated by only integrating over the pion momentum range $15~\text{GeV} < p_\pi < 35~\text{GeV}$ as required by the two signal regions. As already indicated in Table~\ref{tab:ScVePercent}, the SM and LNV cases are approximately equally covered by SR~2 but the LNV distribution in SR~1 is highly suppressed.

\begin{figure}
	\centering
	\includegraphics[width=0.49\textwidth]{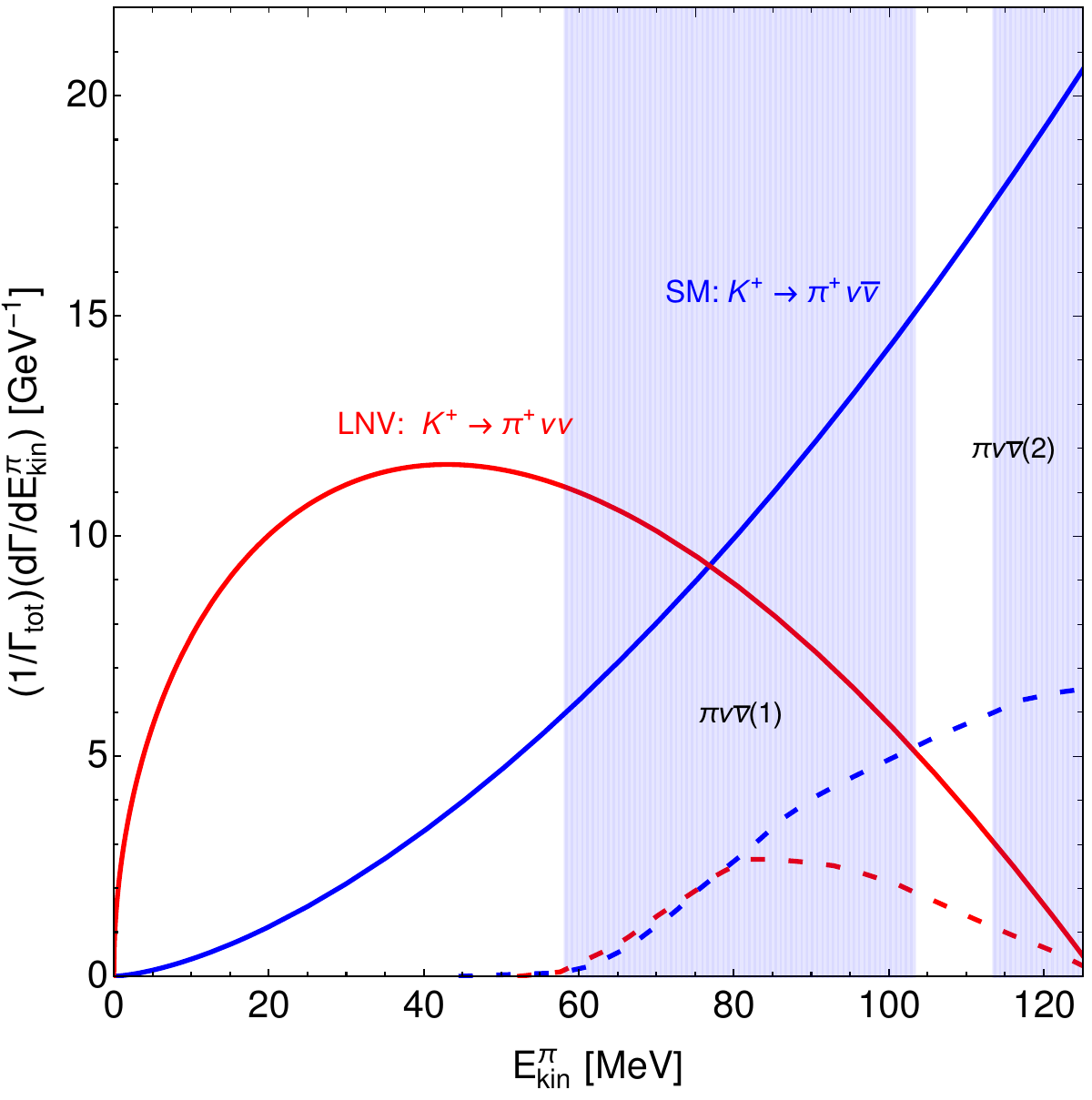}
	\includegraphics[width=0.485\textwidth]{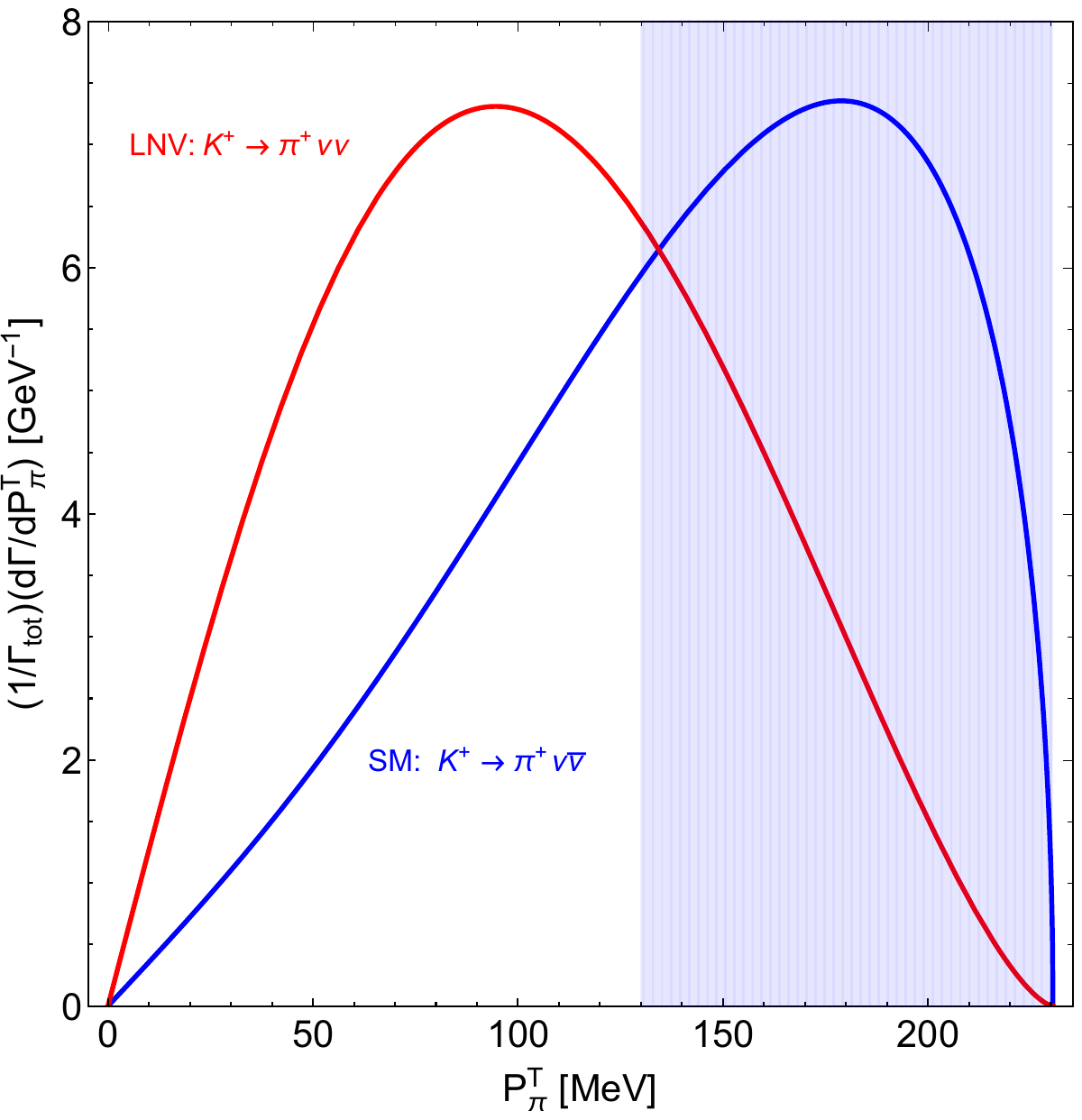}
	\caption{{\bf Left:}
	Normalized differential decay width with respect to the pion kinetic energy in the center of mass frame as applicable in the E949 experiment. The red and blue solid lines correspond to the LNV and SM decays $K^+\to\pi^+\nu\nu$ and $K^+\to\pi^+\nu\bar\nu$, respectively, and the dashed red and blue lines dorrespond to the LNV and SM decays after performing kinematic cuts at the E949 experiment.
	{\bf Right:} The same but showing the normalized differential decay width with respect to the transverse pion momentum relative to the kaon beam with fixed momentum $|\mathbf{p}_K| = 1.4$ GeV, as approximately applicable in the KOTO experiment. In both plots, the shaded areas indicate the signal regions of the E949 and KOTO  experiments, respectively.}
	\label{fig:PionMomentum}
\end{figure}

An analysis as the above can be repeated for the E949 and KOTO experiment, which have different production and detection setups; ideally it should include a dedicated detector simulation to determine the experimental acceptances to the different modes and to determine the constraints or expected sensitivity on the LNV mode, also under the presence of the SM contribution. In Fig.~\ref{fig:PionMomentum} we show the corresponding distributions with respect to the pion momentum in the kaon center-of-mass frame (left) and the pion transverse momentum relative to a boosted kaon direction (right), as approximately applicable in the E949 and KOTO experiments, respectively. The shaded areas again indicate the corresponding signal regions, and in all cases we expect that the experiments are more sensitive to the SM mode with a better coverage within the signal regions. The resulting percentage of the partial width contained within the respective signal regions is shown in Table~\ref{tab:ScVePercent}. For E949, the total SM mode coverage is $\pi\nu\bar\nu(1) + \pi\nu\bar\nu(2) \approx 74\%$ and for the LNV mode it is $\pi\nu\bar\nu(1) + \pi\nu\bar\nu(2) \approx 40\%$. Due to further selection criteria, the acceptance is reduced near the edges of the signal regions leading to a further reduction, in the LNV case. The corresponding distributions for scalar and vector currents, provided in \cite{Artamonov:2009}, after the experimental selection are also shown in Fig.~\ref{fig:PionMomentum}~(left). For KOTO, with the relevant transverse pion momentum distribution shown in Fig.~\ref{fig:PionMomentum}~(right), the selection criteria cover 64\% of SM events but only 30\% of LNV events as shown in Table~\ref{tab:ScVePercent}.

\subsection{Scale of New Physics in $K\to\pi\nu\nu$}
\label{sec:scaleofNP}
We emphasize that we here discuss only the most basic event selection based on the key kinematic properties of the pion. A comprehensive analysis requires a full event simulation with detector effects to determine the sensitivity of the two decay modes in the respective experiment. For example, in Fig.~\ref{fig:PionMomentum}~(right) we omit the dependence on the kaon decay location $Z_\text{vtx}$ on which additional selection criteria apply as discussed in Sec.~\ref{RareKaonExperiment}.

In general, both the SM and LNV mode, will contribute and the total number of observed signal events at a given experiment may be generically expressed as
\begin{align}
	N(K\to\pi\nu\nu) = \Big(\text{BR}(K\to\pi\nu\bar\nu)_\text{SM} A_\text{SM} +
	                         \text{BR}(K\to\pi\nu\nu)_\text{LNV} A_\text{LNV}\Big) N_K.
\end{align}
Here, $N_K$ is the total number of kaons produced and $A_\text{SM}$ and $A_\text{LNV}$ are the overall experimental acceptances in the respective modes. In order to estimate the current limit and future sensitivity we use the branching ratio limits in Sec.~\ref{RareKaonExperiment} and the values in Table~\ref{tab:ScVePercent} to infer the relative acceptance $A_\text{LNV} / A_\text{SM}$.

While E949 provides a limit on scalar currents applicable to the LNV mode, we stress that it is determined using actual experimental data while assuming no SM contribution. We instead use the above description with the relative acceptance $A_\text{LNV} / A_\text{SM} = 0.41$ that we have extracted from the analysis in \cite{Artamonov:2009},
\begin{equation}
\label{eq:E949-limit}
	\text{BR}(K\to\pi\nu\bar\nu)_\text{SM}
	+ \left(\frac{ A_\text{LNV} }{ A_\text{SM}} \right)\times\sum_{i\leq j=1}^3\text{BR}(K\to\pi\nu_i\nu_j)_\text{LNV}	< 3.35\times 10^{-10}
\end{equation}
with $\sum_{i\leq j=1}^3\text{BR}(K\to\pi\nu_i\nu_j)_\text{LNV} = 3\times\text{BR}(K\to\pi\nu_1\nu_1)_\text{LNV}$, to determine the limit on the LNV operator scales. Furthermore, we assume that the operator couples equally to all three SM neutrino species.

\begin{table}
	\centering
	\begin{tabular}{L|LL|LL}
		\specialrule{.2em}{.3em}{.0em}
		\Op & 1/\Lambda^2_{K\to\pi\nu\nu} &  \sum_i\Lambda_{iisd}^\text{E949}\text{ [TeV]} &
		m_{\nu} & \Lambda^{m_\nu}\text{ [TeV]}\\[1mm]
		\hline
		1^{y_d} & \tfrac{v^3}{\Lambda^5} & 2.4 & \tfrac{y_d}{16\pi^2}\tfrac{v^4}{\Lambda^3} & 11.6 \\[1mm]
		3b & \tfrac{v}{\Lambda^3} & 11.5 & \tfrac{y_d}{16\pi^2}\tfrac{v^2}{\Lambda} & 5.2\times 10^{4} \\[1mm]
		3b^{H^2} & {\scriptstyle f(\Lambda)}\tfrac{v }{\Lambda^3} & 5.7 &  \tfrac{y_d}{16\pi^2}\tfrac{v^2 }{\Lambda}{\scriptstyle f(\Lambda)} & 330\\[1mm]
		5 & \tfrac{1}{16\pi^2}\tfrac{ v }{\Lambda^3} & 2.6 & \tfrac{y_d}{\lr{16\pi^2}^2}\tfrac{v^2}{\Lambda} & 330  \\[1mm]
		10 & \tfrac{1}{16\pi^2}\tfrac{y_ev}{\Lambda^3}  & 0.8 & \tfrac{y_ey_d}{\lr{16\pi^2}^2}\tfrac{v^2}{\Lambda} & 9.6\times 10^{-4}\\[1mm]
		11b & \tfrac{1}{16\pi^2}\tfrac{y_dv}{\Lambda^3}  & 0.8 & \tfrac{y_d^2}{\lr{16\pi^2}^2}\tfrac{v^2}{\Lambda} & 8.9\times 10^{-3}\\[1mm]
		14b & \tfrac{1}{16\pi^2}\frac{y_uv}{\Lambda^3}  & 2.9 & \tfrac{y_dy_u}{\lr{16\pi^2}^2}\tfrac{v^2}{\Lambda} & 4.1\times 10^{-3}  \\[1mm]
		66 & {\scriptstyle f(\Lambda)}\tfrac{ v }{\Lambda^3} & 5.1 & \tfrac{y_d}{16\pi^2}\tfrac{v^2 }{\Lambda}{\scriptstyle f(\Lambda)}  & 330 \\[1mm]
		\specialrule{.2em}{.0em}{.3em}
	\end{tabular}
	\caption{Dimension-7 and 9 operators and their effective dimension-6 strength $1/\Lambda_{K\to\pi\nu\nu}^2$ contributing to the rare kaon decay $K\to\pi\nu\nu$. Here, $v$ is the Higgs VEV, with $f(\Lambda) = \left(\tfrac{1}{16\pi^2} + \tfrac{v^2}{\Lambda^2}\right)$, $y_i$ are SM fermion Yukawa couplings and $\Lambda$ is the scale of the operator in question. The constraints are calculated using Eq.~\eqref{eq:E949-limit} with the current E949 limit. We assume that all NP is described by a single operator at a time. The neutrino mass scales $\Lambda^{m_\nu}$ are calculated for first generation Yukawa couplings, assuming a neutrino mass $m_\nu = 0.1$~eV.}
	\label{tab:d7d9op2}
\end{table}
Although several operators could be realised in a complete UV model as exemplified in Sec.~\ref{sec:leptoquarks}, we focus on one operator at a time. Hereby, we do not limit our analysis to $\Op_{3b}$ that is able to trigger $K\to\pi\nu\nu$ at tree level, but consider also operators which could lead to the decay at loop-level. In Table~\ref{tab:d7d9op2}, we present the limit on the NP scale $\sum_i\Lambda_{ii sd}$ for different $\Delta L = 2$ dimension-7 and 9 operators. As expected, the NP scale is most stringently constrained for the dimension-7 operator $\Op_{3b}$ inducing the decay at tree level. Operators of dimension-9 contribute to the rare kaon decay via loops or additional Higgs mass insertions, but are generally more suppressed.

We compare the resulting NP scale with the one required if the studied operator would be the only mechanism to also radiatively generate the neutrino mass. We estimate the corresponding scale as outlined around Eq.~\eqref{eq:numassOp3b} and assume a neutrino mass $m_{\nu} = 0.1$~eV. Except for the operators that require two Yukawa couplings to generate a neutrino mass, the scale of radiative neutrino mass generation is typically higher than the scale of kaon decay, both for first and third generation Yukawa couplings. At face value, this would render any NP in kaon decays with these operators unobservable, as the unobserved Majorana mass of the neutrino provides a more stringent constraint. However, in a UV complete model different contributions could potentially cancel in a non-trivial way such that it is recommended not to constrain oneself by neutrino masses from the start. We discuss such an example in section \ref{sec:leptoquarks}, where we introduce a UV complete realisation including leptoquarks.

\subsection{Fully leptonic LNV meson decays}
\label{sec:leptonic-decays}

Not all operators from Table~\ref{tab:operatorlist} can mediate the rare kaon decay $K\to\pi\nu\nu$ at short range and it is interesting to study the potential of other meson decays. With respect to LNV, in particular meson decays with charged leptons are of interest, as LNV can be directly experimentally observed. While LNV decays such as $K^+ \to \pi^-\ell^+\ell^+$ \cite{Chun:2019nwi}
require at least a dimension-9 operator at short range, the fully leptonic decays $\pi^+\to\mu^+\bar\nu_e$ and $K^+\to\mu^+\bar\nu_e$ \cite{CooperSarkar:1981pb} are possible at dimension-7, although LNV is partially contained in the invisible neutrino, and the $\mu^+\bar\nu_e$ final state implies lepton flavor non-conservation.
These decays can be similarly mediated by the SMEFT operator $\Op_{3b}$,
\begin{equation}
\label{eq:symbreak}
\Op_{3b} = L^\alpha L^ \beta Q^\rho d^cH^\sigma\epsilon_{\alpha\rho}\epsilon_{\beta\sigma} \to h^0 d^c u_L e_L \nu_L.
\end{equation}
However, unlike the decay $K\to\pi\nu\nu$, the fully leptonic LNV meson decays can also be mediated by other dimension-7 operators, such as
\begin{equation}
\label{eq:symbreak2}
\Op_8 = L^\alpha \bar{e}^c\bar{u}^c d^cH^\beta\epsilon_{\alpha\beta}
\to h^0 d^c \bar{u}^c \bar{e}^c \nu_L,
\end{equation}
where the chirality of the final charged lepton and initial up-type quark is different from the case where the decay is mediated by $\Op_{3b}$. The fully leptonic decays can proceed through any operator that is able to mediate the semi-leptonic decays, while the reverse statement does not hold.

In the fully leptonic decays of a pseudo-scalar meson, the hadronic matrix element consists of a parity odd pseudo-scalar meson decaying into vacuum, which is parity-even \cite{Coppola:2018ygv, Dutta:2013qaa}. Hence, the corresponding current in this case has to be parity-odd. Using the standard parametrization, we write the hadronic matrix elements with a vector or axial vector current as
\begin{align}
\label{eq:dconstantsvector}
\bra{0}\bar{q}_2\gamma^\mu q_1\ket{M(p)} &= 0 \nonumber\\
\bra{0}\bar{q}_2\gamma^\mu\gamma_5 q_1\ket{M(p)} &= -if_Mp_\mu,
\end{align}
where $f_M$ is the decay constant of the pseudoscalar meson $M$. From the equation of motion,
\begin{equation}
i\partial_\mu\lr{\bar{q}_2\gamma^\mu\gamma_5 q_1}
= -\lr{m_{q_1}+m_{q_2}}\bar{q_2}\gamma_5q_1,
\end{equation}
with $q_1$ and $q_2$ being the two constituent quarks involved in the interaction, we obtain the corresponding (pseudo-)scalar form factors,
\begin{align}
\label{eq:dconstantsscalar}
\bra{0}\bar{q}_2q_1\ket{M(p)} &= 0, \nonumber\\
\bra{0}\bar{q}_2\gamma_5q_1\ket{M(p)} &= i\frac{m_M^2}{m_{q_1}+m_{q_2}}f_M.
\end{align}
Hence, we can write the matrix element for $M\to\mu^+\bar{\nu}_e$ ($M=K^+, \pi^+$) based on the LNV interaction via $\Op_{3b}$ as
\begin{equation}
\label{eq:melementleptonic}
\iM = i\frac{v}{\Lambda_{ijkn}^3}\lr{m_M^2-m_\mu^2}\frac{m_{M}^2}{m_{q_k} + m_{q_n}}f_M,
\end{equation}
from which the two-body decay width can be easily calculated. We can thus derive the NP scales of LNV operators in a similar fashion to the semi-leptonic case in the previous section,
\begin{align}
	\text{BR}\left(K^+   \to \mu^+ \bar\nu_e\right)
	&= 10^{-3} \left(\frac{2.9~\text{TeV}}{\Lambda_{\mu esd}}\right)^6,\nonumber\\
	\text{BR}\left(\pi^+ \to \mu^+ \bar\nu_e\right)
	&= 10^{-3} \left(\frac{2.0~\text{TeV}}{\Lambda_{\mu eud}}\right)^6.
\end{align}
Here, we specifically focus on the decays $K^+\to\mu^+\bar\nu_e$ and $\pi^+\to\mu^+\bar\nu_e$ as they are experimentally constrained by neutrino oscillation experiment searches \cite{CooperSarkar:1981pb},
\begin{align}
\label{eq:limits-fullyleptonic}
	\text{BR}\left(K^+   \to \mu^+ \bar\nu_e\right) &< 3.3 \times 10^{-3}, \nonumber\\
	\text{BR}\left(\pi^+ \to \mu^+ \bar\nu_e\right) &< 1.5 \times 10^{-3}.
\end{align}
\begin{table}
	\centering
	\begin{tabular}{L|L|L|L|L|L}
		\specialrule{.2em}{.3em}{.0em}
		\Op & 1/\Lambda^2_{M^+\to\ell_i^+\bar\nu_j}& \Lambda_{\mu e us}\text{ [TeV]}& \Lambda_{\mu e ud}\text{ [TeV]}& m_\nu & \Lambda^{m_\nu}\text{ [TeV]}\\[1mm]
		\hline
		3a & \tfrac{v}{\Lambda^3} & 2.2 & 1.7 & \tfrac{y_d g^2}{\lr{16\pi^2}^2}\tfrac{v^2}{\Lambda} & 69 \\[1mm]
		3a^{H^2} & {\scriptstyle f(\Lambda)}\tfrac{v}{\Lambda^3} & 1.3 & 1.1 & \tfrac{y_d g^2}{\lr{16\pi^2}^2}\tfrac{v^2}{\Lambda}{\scriptstyle f(\Lambda)} & 0.4 \\[1mm]
		4a & \tfrac{v}{\Lambda^3} & 2.2 & 1.7 & \frac{y_u}{16\pi^2}\tfrac{v^2}{\Lambda} & 2.4\times10^4 \\[1mm]
		4a^{H^2}& {\scriptstyle f(\Lambda)}\tfrac{v}{\Lambda^3} & 1.3 & 1.1 & \frac{y_u}{16\pi^2}\tfrac{v^2}{\Lambda}{\scriptstyle f(\Lambda)} & 150 \\[1mm]
		4b^\dagger & \tfrac{v}{\Lambda^3} & 2.2 & 1.7 & \tfrac{y_u g^2}{\lr{16\pi^2}^2}\tfrac{v^2}{\Lambda} & 33\\[1mm]
		4b^{\dagger H^2} & {\scriptstyle f(\Lambda)}\tfrac{v}{\Lambda^3} & 1.3 & 1.1 & \tfrac{y_u g^2}{\lr{16\pi^2}^3}\tfrac{v^2}{\Lambda} & 0.2\\[1mm]
		6 & {\scriptstyle f(\Lambda)}\tfrac{v}{\Lambda^3} & 1.3 & 1.1 & \tfrac{y_u}{\lr{16\pi^2}^2}\tfrac{v^2}{\Lambda} & 150  \\[1mm]
		7 & \tfrac{v^3}{\Lambda^5} & 0.8 & 0.7 & \tfrac{y_eg^2}{\lr{16\pi^2}^2}\tfrac{v^2}{\Lambda}{\scriptstyle f(\Lambda)} & 0.6 \\[1mm]
		8  & \tfrac{v}{\Lambda^3} & 2.2 & 1.7 & \tfrac{y_ey_dy_ug^2}{\lr{16\pi^2}^2}\tfrac{v^4}{\Lambda^3} & 4.3\times 10^{-4} \\[1mm]
		8^{H^2} & {\scriptstyle f(\Lambda)}\tfrac{ v}{\Lambda^3} & 1.3 & 1.1 & \tfrac{y_ey_dy_ug^2}{\lr{16\pi^2}^2}\tfrac{v^4}{\Lambda^3}{\scriptstyle f(\Lambda)} & 7.9\times 10^{-5} \\[1mm]
		11a & \tfrac{1}{16\pi^2}\tfrac{y_dv}{\Lambda^3} & 0.2 & 0.1 & \frac{y_d^2g^2}{\lr{16\pi^2}^3}\tfrac{v^2}{\Lambda} & 1.2\times 10^{-5}\\[1mm]
		12a & \tfrac{1}{16\pi^2}\tfrac{y_uv}{\Lambda^3} & 0.6 & 0.5 & \tfrac{y_u^2}{\lr{16\pi^2}^2}\tfrac{v^2}{\Lambda} & 1.9\times 10^{-3} \\[1mm]
		12b^* & \tfrac{1}{16\pi^2}\tfrac{y_uv}{\Lambda^3} & 0.7 & 0.6 & \frac{y_u^2g^2}{\lr{16\pi^2}^3}\tfrac{v^2}{\Lambda} & 2.6\times10^{-6}\\[1mm]
		13 & \tfrac{1}{16\pi^2}\tfrac{y_ev}{\Lambda^3}  & 0.2 & 0.2 & \tfrac{y_ey_u}{\lr{16\pi^2}^2}\tfrac{v^2}{\Lambda} & 4.5\times 10^{-4}\\[1mm]
		14a & \tfrac{1}{16\pi^2}\frac{\lr{y_u + y_d}v}{\Lambda^3} & 0.6 & 0.5 & \tfrac{ y_u y_d g^2}{\lr{16\pi^2}^3}\tfrac{v^2}{\Lambda} & 5.6\times10^{-6}\\[1mm]
		16 & \tfrac{1}{16\pi^2}\tfrac{y_ev}{\Lambda^3} & 0.1 & 0.1 & \tfrac{y_dy_ug^4}{\lr{16\pi^2}^4}\tfrac{v^2}{\Lambda} & 7.4\times10^{-9}\\[1mm]
		19 & \tfrac{1}{16\pi^2}\tfrac{y_dv}{\Lambda^3} & 0.1 & 0.1 & \tfrac{y_ey_uy_d^2g^2}{\lr{16\pi^2}^3}\tfrac{v^4}{\Lambda^3} & 2.4\times10^{-6} \\[1mm]
		20 & \tfrac{1}{16\pi^2}\tfrac{y_uv}{\Lambda^3} & 0.5 & 0.4 & \tfrac{y_ey_u^2y_dg^2}{\lr{16\pi^2}^3}\tfrac{v^4}{\Lambda^3} & 1.8\times10^{-6}\\[1mm]
		\specialrule{.2em}{.0em}{.3em}
	\end{tabular}
	\caption{Dimension-7 and 9 operators and the effective dimension-6 operators strength contributing to the LNV fully leptonic charged pion and kaon decays. Here $v$ is the Higgs VEV, with $f(\Lambda) = \left(\tfrac{1}{16\pi^2} + \tfrac{v^2}{\Lambda^2}\right)$, $y_i$ are the SM fermion Yukawa couplings, $\Lambda$ is the NP scale of the operator in question, and $g$ is the weak coupling constant. We only list operators that do not also contribute to the rare kaon decay $K\to\pi\nu\nu$. The limits on the scales $\Lambda^{K^+,\pi^+}_{\mu e kn}$ are determined using the experimental constraints in Eq.~\eqref{eq:limits-fullyleptonic}. The neutrino mass scales $\Lambda^{m_\nu}$ are calculated for first generation Yukawa couplings, assuming a diagonal CKM matrix and neutrino mass $m_{\nu} = 0.1$ eV.}
	\label{tab:semileptonicd7d9op2}
\end{table}
Other dimension-7 and 9 LNV operators can similarly be constrained and the results are shown in Table~\ref{tab:semileptonicd7d9op2}. Note that we do not include the operators that we studied in the context of semi-leptonic decays in Table~\ref{tab:d7d9op2}, since we expect the semi-leptonic decays to put more stringent constraints on these operators than the fully leptonic decays. This can indeed be seen by comparing for example the limit on the NP scales of operators $\Op_{3b}$ from Table~\ref{tab:d7d9op2} ($11.5$~TeV) and $\Op_{3a}$ from Table~\ref{tab:semileptonicd7d9op2} ($2.4$~TeV). 
Constrained by the fully leptonic decays, operator $\Op_{3b}$ would similarly lead to a lower limit of $2.4$~TeV. Therefore, the operators that are able to mediate semi-leptonic meson decays are not considered for the fully leptonic analysis. As in the semi-leptonic case, the dimension-7 operators provide more stringent limits than the dimension-9 ones. Similarly, the radiative neutrino mass generation scales are generally higher than the ones from meson decays, except when multiple first generation Yukawa couplings are involved. In the LNV decay of a $\pi^+$, the quark flavour content is the same as in $0\nu\beta\beta$ decay, and the corresponding operator would, in this respect, generally get more stringent contraints on the NP scale from $0\nu\beta\beta$ decay than from the $\pi^+$ decay. However, specifically in the LNV decay $\pi^+\to\mu^+\bar\nu_e$, the lepton flavour content is dissimilar from that of $0\nu\beta\beta$ decay, and the corresponding NP scale could differ in a flavour non-blind UV complete scenario. 

\subsection{Overview of LNV probes}
Apart from the decays of kaons and pions discussed above, the different operators of Table \ref{tab:operatorlist} could be observed in LNV processes such as $\mu^-$ to $e^+$ conversion, $0\nu\beta\beta$ decay, and other LNV meson decays such as that of B-mesons. We do not attempt to give a full account of all processes and modes. In Table~\ref{tab:d7mesondecay2}, we instead compare the NP scales for these processes determined by the operator constraining this observable the most. By far, the highest limit on the NP scale comes from $0\nu\beta\beta$ decay \cite{Deppisch:2018}. However, this observable is only sensitive to LNV in electrons, as well as only to first generation quarks.
\begin{table}
	\centering
	\begin{tabular}{l|l|l|l|l}
		\specialrule{.2em}{.3em}{.0em}
		Process & Experimental limit & $\Op$ &$\Lambda^{\mathrm{NP}}_{ijkn}$  [TeV] & $\hat{\lambda}$ [TeV]\\[1mm]
		\hline
		$K^+\to\pi^+\nu\nu$      & $\text{BR}_{\text{future}}^{\text{NA62}} < 1.11 \times 10^{-10}$  &
		$\Op_{3b}$ & $\sum_i\Lambda_{ii sd} > 19.6$ & 0.213 \\
		$K^+\to\pi^+\nu\nu$      & $\text{BR}_{\text{current}}^{\text{NA62}} < 1.78 \times 10^{-10}$ \cite{CortinaGil:2020vlo} &
		$\Op_{3b}$ & $\sum_i\Lambda_{ii sd} > 17.2$ & 0.196 \\
		$K_L\to\pi^0\nu\nu$      & $\text{BR}_{\text{current}}^{\text{KOTO}} < 3.0 \times 10^{-9}$ \cite{Masuda:2015eta} &
		$\Op_{3b}$ & $\sum_i\Lambda_{ii sd} > 12.3$ & 0.178 \\[.5mm]
		\specialrule{.12em}{.0em}{.0em}
		$B^+\to\pi^+\nu\nu$      & $\text{BR} < 1.4\times 10^{-5}$ \cite{PDG:2018} &
		$\Op_{3b}$ & $\sum_i\Lambda_{ii bd} > 1.4$ & 0.174 \\
		$B^+\to K^+\nu\nu$       & $\text{BR} < 1.6\times 10^{-5}$ \cite{PDG:2018} &
		$\Op_{3b}$ & $\sum_i\Lambda_{ii bs} > 1.4$ & 0.174 \\
		$B^0\to\pi^0\nu\nu$      & $\text{BR} < 9 \times 10^{-6}$ \cite{PDG:2018} &
		$\Op_{3b}$ & $\sum_i\Lambda_{ii bd} > 1.5$ & 0.174 \\
		$B^0\to K^0\nu\nu$       & $\text{BR} < 2.6 \times 10^{-5}$ \cite{PDG:2018} &
		$\Op_{3b}$ & $\sum_i\Lambda_{ii bs} > 1.3$ & 0.174 \\
		$K^+\to\mu^+\bar\nu_e$   & $\text{BR} < 3.3 \times 10^{-3}$ \cite{CooperSarkar:1981pb} &
		$\Op_{3a}$ & $\Lambda_{\mu esu} > 2.4$ & 0.174 \\
		$\pi^+\to\mu^+\bar\nu_e$ & $ \text{BR} < 1.5 \times 10^{-3}$ \cite{CooperSarkar:1981pb} &
		$\Op_{3a}$ & $\Lambda _{\mu eud} > 1.9$ & 0.174 \\
		$\pi^0\to\nu\nu$         & $\text{BR} < 2.9 \times 10^{-13}$ \cite{Lam:1991bm} &
		$\Op_{3b}$ & $\Lambda _{\nu\nu ud} > 3.4$ & 0.174 \\
		$0\nu\beta\beta $        & $T_{1/2}^{^{136}\text{Xe}} \leq 1.07\times10^{26}\text{ yrs}$ \cite{KamLAND:2016} &
		$\Op_{3b}$ & $\Lambda_{eeud} > 330$ & 3.5 \\
		$\mu^-\to e^+ $          & $R_{\mu^-e^+}^{\text{Ti}} < 1.7\times 10^{-12}$ \cite{Kaulard:1998rb} &
		$\Op_{14b}$ & $\Lambda_{\mu eud} > 0.01$ & 0.174 \\
		\specialrule{.2em}{.0em}{.3em}
	\end{tabular}
	\caption{Selected limits on LNV operator scales from different experimental constraints. For NA62 and KOTO we have estimated the sensitivity to scalar currents using the theoretical decay widths in the fiducial phase space of the respective experiment. The NP scales for $0\nu\beta\beta$ decay and $\mu^-\rightarrow e^+$ conversion are calculated in \cite{Deppisch:2018} and \cite{Berryman:2016slh}, respectively. The quantity $\hat\lambda$ denotes the temperature above which the washout of lepton asymmetry due to the given operator and scale is highly effective.}
	\label{tab:d7mesondecay2}
\end{table}
In the context of the rare decay $K\to\pi\nu\nu$, we have already discussed the NA62, KOTO and E949 experiments, where we have focussed on the latter to set bounds on the scalar mode induced by the LNV operator. We here also estimate the limits on $K^+\to\pi^+\nu\nu$ and $K_L\to\pi^0\nu\nu$ from NA62 and KOTO. Because these experiments do not attempt to constrain scalar currents, we estimate their acceptance using the theoretical decay width integrated over the fiducial phase space of the respective experiment as given in Table~\ref{tab:ScVePercent}. The limit on the LNV operator scale is then determined analogous to Eq.~\eqref{eq:E949-limit}, with $A_\text{LNV}/A_\text{SM} = 0.23/0.15$ for NA62 and $A_\text{LNV}/A_\text{SM} = 0.64/0.30$ for KOTO.

We have already discussed fully leptonic LNV decays of pions, an example of which is listed in Table~~\ref{tab:d7mesondecay2}. Another interesting mode is $\pi^0\to\nu\nu$ where the strongest limit originates from Big Bang Nucleosynthesis for Dirac neutrinos. Since we are considering Majorana neutrinos in our analysis, this limit is mainly interesting for comparison in a broader perspective.
A less stringent limit of BR~$< 2.7\times 10^{-7}$ at 90\% CL \cite{Artamonov:2005cu}, coming from colliders, and not necessarily relying on Dirac neutrinos, would lead to a lower scale $\Lambda _{\mu eud} > 0.3$~TeV. Constraints coming from the LNV decays of the bottom mesons $B^+$ and $B^0$ are generally less stringent than those coming from kaons, but are nevertheless interesting, as the bottom mesons probe couplings to the third generation quarks, as well as second or first generation, depending on whether the final state contains a kaon or pion. For strange quarks, the most constraining observable is indeed the rare kaon decay $K^+\to\pi^+\nu\nu$ based on the experimental limit from the E949 experiment and is expected to be improved by NA62 in the near future.

In addition to the experimental constraints, we show in Table~\ref{tab:d7mesondecay2} the temperature $\hat\lambda$ below which the washout of lepton number would cease to be efficient based on the scale that was derived from the corresponding observable. This was discussed in Section~\ref{leptogenesis}. A lepton and thus baryon asymmetry (if sphalerons are in thermal equilibrium) will be effectively washed out. This implies that an observation of one of the listed processes will immediately lead to tension with baryogenesis scenarios at higher scales. In order to be fully conclusive, all flavours have to be equilibrated or lepton flavour violation around the same NP scales should be observed.

%% file: leptoquarks.tex
\section{Example for a possible UV completion: Leptoquarks}\label{sec:leptoquarks}
\begin{figure}[t]
\centering
    \includegraphics[width=0.49\textwidth]{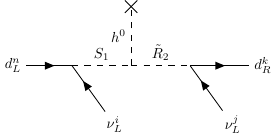}
    \includegraphics[width=0.49\textwidth]{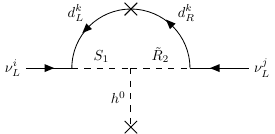}\\
    \includegraphics[width=0.49\textwidth]{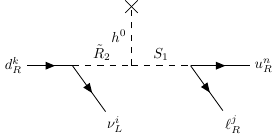}
    \includegraphics[width=0.49\textwidth]{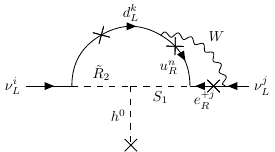}
    \caption{Contributions of the simplified $\tilde{R}_2$-$S_1$ leptoquark model via the interactions in Eq.~\eqref{eq:mannellagrangian} to meson decays with one or two neutrinos in the final state (left) and to the neutrino mass (right). The top diagrams correspond to contributions arising from $\Op_{3b}$, whereas the bottom diagrams correspond to $\Op_{8}$.}
    \label{fig:mnudiagram}
\end{figure}

In the following, we discuss a possible UV completion giving rise to the operator $\Op_{3b}$ that we studied in the context of rare kaon decay $K\to\pi\nu\nu$ and $\Op_{8}$ that we analyzed in the context of the fully leptonic kaon decay modes. Following the conventions of \cite{Dorsner:2016}, we can generate $\Op_{3b}$ by a pair of scalar leptoquarks, namely via the combination of the scalar leptoquark $\tilde{R}_2$ with either $S_1$ or $S_3$. Under the SM gauge group $SU(3)_c\times SU(2)_L \times U(1)_Y$, they have the following representations
\begin{equation}
\label{eq:LQfields}
\begin{aligned}
	&\tilde{R}_2 \in \lr{3, \; 2, \; 1/6},\\
	&S_1 \in \lr{\bar{3}, \; 1, \; 1/3},\\
	&S_3 \in \lr{\bar{3}, \; 3, \; 1/3}\,.
\end{aligned}
\end{equation}
Scalar leptoquarks have also been studied in their relation to the $R_{K^{(*)}}$ \cite{Bifani:2018zmi} anomaly \cite{Kosnik:2012dj,Fajfer:2018bfj,Cata:2019,Descotes-Genon:2020buf}. For simplicity we focus on one combination of leptoquarks, $\tilde{R}_2$ and $S_1$, and extend the SM Lagrangian correspondingly~\cite{Cata:2019},
\begin{equation}
\label{eq:fullLQlagrangian}
\begin{aligned}
    \mathcal{L} &= \mathcal{L}_{\text{SM}} - \tilde{R}_2^{\dagger\alpha}(\Box + m_{\tilde{R}_2}^2)\tilde{R}_{2\alpha} - S_1^*(\Box + m_{S_1}^2)S_1\\
    &+ \mu S_1 H^{\dagger\alpha}\tilde{R}_{2\alpha} - g_1^{ik}\bar{L}_{i\alpha} i\sigma_2^{\alpha\beta}\tilde{R}_{2\beta}^{*}\overline{d}^c_{k} - g_2^{jn}Q^{\alpha}_nL_j^\beta\epsilon_{\alpha\beta}S_1\\
    &-g_3^{jn}\bar{u}_{n}^ce_jS_1 + \text{h.c.}\,.
    \end{aligned}
\end{equation}
Here, Roman letters indicate flavour indices and Greek letters $SU(2)_L$ indices, and $\sigma_2$ stands for the second Pauli matrix. Furthermore, $u$, $d$, and $e$ are the usual $SU(2)_L$ singlets. The leptoquarks $\tilde{R}_2$ and $S_1$ both carry lepton number $L = -1$, but differ in their baryon number, with $\tilde{R}_2$ carrying $B=1/3$ and $S_1$ baryon number $B=-1/3$ \cite{Dorsner:2016}.
The coupling between the leptoquarks and the SM Higgs features an LNV vertex with $\mu$ being a dimensionful coupling which we assume to be real. In order to match it with our previously introduced effective operators, we integrate out the heavy leptoquarks (assuming $m_{\tilde{R}_2}$, $m_{S_1}$  $\gg$ $\Lambda_{\text{EW}}$). We arrive at two effective contributions \cite{Cata:2019}: Firstly, we obtain a dimension-6 lepton number conserving part,
\begin{equation}
\begin{aligned}
\label{eq:6dimmannellagrangian}
    \mathcal{L}_{6D} &=
				-\frac{g^{ik}_{1}g^{*jn}_{1}}{2m_{\tilde{R}_2}^2}\bar{d}_n\gamma^\mu d_k\bar{L}^\alpha_i\gamma_\mu L_{j\alpha}+\frac{g^{ik}_{2}g^{*jn}_{2}}{2m_{S_1}^2}\epsilon_{\alpha\beta}\epsilon_{\rho\sigma}\bar{Q}_n^\alpha\gamma^\mu Q_k^\rho\bar{L}^\beta_i\gamma_\mu L_{j}^\sigma\\
				&+\frac{g^{ik}_{3}g^{*jn}_{3}}{2m_{S_1}^2}\bar{u}_n\gamma^\mu u_k\bar{e}_i\gamma_\mu e_{j}-\frac{g^{ik}_{2}g^{*jn}_{3}}{2m_{S_1}^2}\epsilon_{\alpha\beta}\lr{\bar{u}_n Q_k^\alpha\bar{e}_i L_{j\beta} - \frac{1}{4}\bar{u}_n\sigma^{\mu\nu} Q_k^\alpha\bar{e}_i\sigma_{\mu\nu} L_{j}^\beta},
				\end{aligned}
\end{equation}
with $\sigma^{\mu\nu} = \frac{i}{2}\left[\gamma^\mu,\gamma^\nu\right]$, where the second term resembles the SM contribution leading to a vector current. Secondly, we acquire
a dimension-7 lepton number violating part,
\begin{equation}
\label{eq:mannellagrangian}
    \mathcal{L}_{7D} = \frac{\mu g^{ik}_{1}g^{jn}_{2}}{m_{\tilde{R}_2}^2 m_{S_1}^2}L^\alpha_i H^\beta{d}^c_kQ_n^\mu L^\nu_j\epsilon_{\alpha\beta}\epsilon_{\mu\nu} + \frac{\mu g^{ik}_{1}g^{jn}_{3}}{m_{\tilde{R}_2}^2 m_{S_1}^2}L^\alpha_i H^\beta{d}^c_ku_n^c e^c_j\epsilon_{\alpha\beta}.
\end{equation}
We can identify the first term in Eq.~\eqref{eq:mannellagrangian} with operator $\Op_{3b}$ and the second term with operator $\Op_8$. As discussed before, both terms will generate radiative neutrino masses and similarly will lead to meson decays if all lepton and quark generations couple similarly to the leptoquarks. We show the corresponding diagrams in Fig.~\ref{fig:mnudiagram}.
As discussed in Section~\ref{sec:scaleofNP} and displayed in Tab.~\ref{tab:d7d9op2}, the experimental bounds on the neutrino mass constrain the scale of the corresponding operators more stringently than arising from kaon decays. Hence, an a LNV contribution to rare kaon decays would imply some non-trivial flavour pattern contributing to the neutrino mass generation. For example, flavour-specific couplings between the leptoquarks and quarks could lead to a suppression within the neutrino mass contributions while still contributing to the rare kaon decays.
In order to demonstrate such a situation, we relate the effective operator scale $\Lambda_{ijkn}$ with the model parameters of the simplified $\tilde{R}_2$-$S_1$ leptoquark model with the coupling constants of the two leptoquarks being matrices in flavour space,
\begin{equation}
\label{eq:thefinalkaonscale}
    \frac{1}{\Lambda_{ijkn}^3} \equiv \frac{\mu g_{1}^{ik} g_{2/3}^{jn} }{m_{\tilde{R}_2}^2 m_{S_1}^2}.
\end{equation}
According to the results given Table~\ref{tab:d7d9op2} for $\Op_{3b}$\footnote{The constraint on the neutrino mass for $\Op_{8}$ from Table~\ref{tab:semileptonicd7d9op2}, corresponding to the coupling of $S_1$ to $SU(2)_L$ singlet fields, is less stringent than that of $\Op_{3b}$, corresponding to the coupling of $S_1$ to $SU(2)_L$ doublet fields.}, we realize that the constraint coming from the radiative neutrino mass generation is stronger than the one from kaon decay under the assumption of flavor democratic couplings. Hence, an observation of the rare kaon decay would either point us towards a lepton conserving contribution only or a flavor specific coupling of the leptoquarks.

\begin{figure}[t]
\centering
    \includegraphics[width=0.49\textwidth]{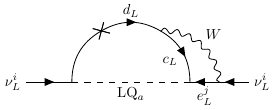}
    \caption{Radiative neutrino mass contribution in case of flavour specific couplings. Here $\text{LQ}_a$ is a mass eigenstate of the $\tilde{R}_2$-$S_1$ leptoquark system. The leptoquark $\tilde{R}_2$ is assumed to couple to first generation down-type quarks only, and $S_1$ interacts with second generation quarks. }
    \label{fig:mnu2loopdiagramQ}
\end{figure}

For example, if $\tilde{R}_2$ couples only to first generation right-handed down-type quarks, and $S_1$ only to second generation left-handed quarks (the first term in Eq.~\eqref{eq:mannellagrangian}), the rare kaon decay would be enhanced while the neutrino mass would not be generated at 1-loop but only at 2-loop level, as demonstrated in Fig.~\ref{fig:mnu2loopdiagramQ}. This two loop mass can be expressed as
\begin{equation}
\label{eq:thefinalnumass}
	(m_\nu)_{i} = \sum_j\frac{3\sin(2\theta)g^2V_{cd}\tilde{g}_1^{id}\tilde{g}_2^{jc}U_{ji}}{512\pi^4}m_dI(m_{\text{LQ}_1}^2,m_{\text{LQ}_2}^2,m_W^2),
\end{equation}
where the index $i$ is not summed over. Here $g$ is the weak coupling constant, $V_{cd}$ is a CKM matrix element, $U_{ji}$ is a PMNS matrix element, $m_d$ is the mass of the first generation down-type quark, $m_W$ is the mass of the $W$ boson, and $\theta$ is the mixing angle between $S_1$ and one of the components of $\tilde{R}_2$, which diagonalizes the mass matrix
\begin{equation}
M^2=\begin{pmatrix}
m_{\tilde{R}_2}^2 & \mu v\\
\mu v & m_{S_1}^2
\end{pmatrix},
\end{equation}
leading to the leptoquark mass eigenvalues $m_{\text{LQ}_1}$ and $m_{\text{LQ}_2}$. The angle $\theta$ is given by \cite{AristizabalSierra:2007nf,Babu:2010vp,Dorsner:2017wwn}
\begin{equation}
\tan(2\theta)=\frac{2\mu v}{m_{\tilde{R}_2}^2-m_{S_1}^2}.
\end{equation}
Furthermore, $\tilde{g}_1^{id}$ and $\tilde{g}_2^{jc}$ are leptoquark couplings in the SM fermion mass basis, which are given by
\begin{equation}
	\tilde{g}_1^{ik}=\sum_{\alpha}g_1^{\alpha k}U^{i\alpha},\quad\tilde{g}_2^{jn}=\sum_{\alpha}g_2^{j\alpha}V^{\alpha n}.
\end{equation}
In the approximation that the quarks and charged leptons are massless, the loop function $I(m_{\text{LQ}_1}^2,m_{\text{LQ}_2}^2,m_W^2)$ is given by \cite{Babu:2010vp}
\begin{equation}
\begin{aligned}
I(m_{\text{LQ}_1}^2,m_{\text{LQ}_2}^2,m_W^2) &\approx \lr{1-\frac{m_{\text{LQ}_1}^2}{m_{\text{LQ}_2}^2}}\times\\
&\Bigg[1+\frac{\pi^2}{3}+\frac{m_{\text{LQ}_1}^2\log\frac{m_{\text{LQ}_2}^2}{m_W^2}-m_{\text{LQ}_2}^2\log\frac{m_{\text{LQ}_1}^2}{m_W^2}}{m_{\text{LQ}_2}^2-m_{\text{LQ}_1}^2}+\\
&\frac{1}{2}\frac{m_{\text{LQ}_1}^2\lr{\log\frac{m_{\text{LQ}_2}^2}{m_W^2}}^2-m_{\text{LQ}_2}^2\lr{\log\frac{m_{\text{LQ}_1}^2}{m_W^2}}^2}{m_{\text{LQ}_2}^2-m_{\text{LQ}_1}^2}\Bigg],
\end{aligned}
\end{equation}
where $m_{\text{LQ}_1}$ is the lighter of two the mass eigenstates.
As an example, we choose leptoquark masses $m_{\tilde{R}_2} = 4$ TeV and $m_{S_1} = 2$ TeV close to the experimental discovery reach of the LHC \cite{Khachatryan:2016} and the dimensionful coupling as $\mu = 10$ GeV, in agreement with current experimental bounds on the electroweak $\rho$ parameter \cite{PDG:2018}. Furthermore, we assume $\tilde{g}_1^{id} = \tilde{g}_2^{jc} = 1$, with all other leptoquark couplings in the mass basis being zero. With the PMNS matrix assumed to be the identity matrix, this results approximately in $\tilde{g}_2^{jc} = {g}_2^{js}$. From Eq.~\eqref{eq:thefinalkaonscale} and Eq.~\eqref{eq:thefinalnumass}, we then find a radiative neutrino mass of $(m_\nu)_{i} \approx 0.08$ eV, and a NP scale in rare kaon decays of $\Lambda_{sdij} \approx 18.6$ TeV. In this benchmark scenario, the NP scale of LNV rare kaon decays is close to the experimental limit (see Table \ref{tab:d7mesondecay2}), while the radiatively generated neutrino mass is small. Although a specific choice of parameters has been made, this example demonstrates that non-trivial flavour patterns in a UV complete model can in principle lead to a suppression of the neutrino mass generation.

\begin{figure}[t]
\centering
    \includegraphics[width=0.49\textwidth]{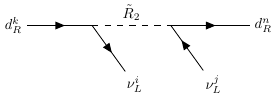}
    \includegraphics[width=0.49\textwidth]{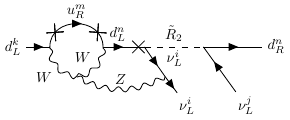}\\
    \includegraphics[width=0.49\textwidth]{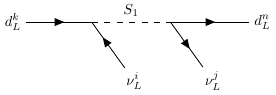}
    \includegraphics[width=0.49\textwidth]{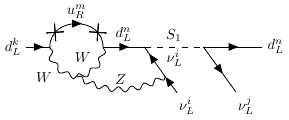}
    \caption{Meson decay at dimension 6 with (right) and without (left) flavour specific couplings. The top diagrams correspond to the first term in Eq.~\eqref{eq:6dimmannellagrangian}, the bottom diagrams correspond to the second term.}
    \label{fig:D6LQdiagram}
\end{figure}

As pointed out before, the Lagrangian in Eq.~\eqref{eq:6dimmannellagrangian} also features a lepton number conserving dimension-6 contribution that similarly mediates rare kaon decay $K\rightarrow\pi\nu\bar{\nu}$ with the mediator being  either a $\tilde{R}_{2}$ or $S_1$ leptoquark as shown in Fig.~\ref{fig:D6LQdiagram} (left).
However, in case of a flavour specific pattern, these operators do not contribute at tree level Fig.~\ref{fig:D6LQdiagram} (left), but only at loop level as depicted in Fig.~\ref{fig:D6LQdiagram} (right), leading to a GIM suppression of the SM-like contribution.

Currently, the most stringent experimental bound on the mass of first generation scalar leptoquarks is provided by the CMS collaboration \cite{Khachatryan:2016}, $m_{\mathrm{LQ}} > 1755$ GeV at 95\% CL. This limit is still weaker than from considerations of meson decays. This shows that leptoquarks are an interesting UV completion featuring an interplay of various constraints of different types of experiments.

%% file: conclusions.tex
\section{Conclusions}\label{sec:summary}
The observation of a LNV process would have far reaching consequences in our understanding of particle physics. Most importantly it would point towards a Majorana nature of neutrinos and it would have implications on the viability of leptogenesis scenarios. While $0\nu\beta\beta$ decay is the crucial probe to test the Majorana nature of neutrinos and LNV interactions in general, as a nuclear process it is limited to first generation leptons and quarks only. Hence, it is worthwhile to explore other LNV probes.

In this work, we have considered the possibility of LNV interactions in the rare meson decay $K\to\pi\nu\nu$. The GIM and loop-suppressed SM branching ratio of this decay can be determined very precisely due to its comparably tiny hadronic uncertainties. With the NA62 experiment aiming for a 10\% sensitivity on the SM branching ratio in the future, it is consequently an excellent probe for new physics. In the SM, the kaon is expected to decay into a neutrino and an anti-neutrino. In case of Majorana neutrinos, however, additional LNV interactions can be present, leading to an emission of two neutrinos or anti-neutrinos. Since the experiments are not able to directly detect the neutrinos, both options should be considered. While their impact on the operator scales have already been explored in Ref.~\cite{Li:2019fhz}, we investigated their implications with respect to their different leptonic current structure.

Within a SMEFT, we identified one dimension-7 LNV operator $h_0 d^c s_L \nu_L \nu_L$ that can generate $K\to\pi\nu\nu$ at tree level leading to a scalar leptonic current. This is in contrast to the usual SM contribution that features a leptonic vector current. We demonstrated that this results in different kinematic distributions of the pion momentum for both cases. Contrasting this with the signal regions of NA62, we generally expect more events in the lower pion momentum region for the SM case but an order of magnitude fewer events with an LNV contribution. Based on the expected kinematic distribution only, we expect NA62 to be less sensitive in the LNV case, though this will still be subject to the specific detector acceptance in the corresponding regions. However, the expected difference in the phase space distribution with respect to the SM and LNV contribution could be used to probe a possible LNV admixture. Similar conclusions can be drawn for the E949 and KOTO experiment, though global sensitivity to the LNV contribution might be reduced.

Moreover, we can set limits on the scale of the operator based on the current upper limits on the branching ratios. While the strongest limits were just recently improved by the NA62 experiment to $\text{BR}(K^+\to\pi^+\nu\bar\nu) < 1.78\times 10^{-10}$ \cite{NA62talkICHEP}, the E949 experiment constrained it previously to $\text{BR}(K^+\to\pi^+\nu\bar\nu) < 3.35\times 10^{-10}$ \cite{Artamonov:2009}. As only the E949 experiment provides us with the relative acceptance between a SM-like vector current and a scalar current, we were able to set a lower limit on the corresponding dimension-7 LNV operator scale $\sum_i \Lambda^\text{E949}_{iisd} > 11.5$~TeV. In contrast, the NA62 experiment provides a limit on the SM-like vector current only. Taking this limit, we obtain an seemingly stronger limit on the scale of $\sum_i \Lambda^\text{NA62}_{iisd} > 17.2$~TeV, but this is expected to be weaker for the scalar current case due to the reduced experimental acceptance. Similar reasoning applies for the results of the KOTO experiment. With the limits given for a vector current, we arrive at $\sum_i \Lambda^\text{KOTO}_{iisd} > 12.3$~TeV. In future, NA62 and KOTO are expected to improve their reach such that e.g. a scale $\sum_i \Lambda^\text{NA62,~future}_{iisd} \approx 19.6$~TeV can be probed. Rare kaon decays are thus able to probe very high scales of LNV physics.

We stress the importance of dedicated limits on a scalar current contribution by the NA62 and KOTO experiments in the future, as this will be very useful for studying LNV scenarios and their far reaching consequences. By comparing the relative contributions to the different signal regions, one might be able to draw conclusions on the existence of an additional LNV interaction. For example, while we theoretically expect a ratio of events of $\approx 0.35$ between the lower and higher momentum signal regions of NA62 for a SM contribution only, the event ratio for the LNV scenario is much smaller $\approx 0.02$.

Despite focussing on LNV operators in SMEFT, i.e. where only light active Majorana neutrinos are present, we must emphasise that the presence of LNV cannot be strictly proven using the process $K\to\pi\nu\nu$ as the two neutrinos (anti-neutrinos) are not experimentally observable. While the kaon decay distribution is different from that in the SM due to the scalar nature of the currents involved, this behaviour can also emerge in an EFT that includes sterile neutrino states in addition to the SM particle content. Such a framework can encompass scenarios where (i) the light active neutrinos are Dirac fermions but participate in interactions beyond the SM and (ii) there are additional sterile neutrinos (light enough to be produced in the decay) with exotic interactions. In such scenarios, total lepton number may be conserved depending on the nature of the sterile states.

For completeness, we have not only focused in our study on the semi-leptonic rare kaon decays $K\to\pi\nu\nu$, but we discussed also a possible LNV contribution to fully leptonic meson decays such as $K^+\to \mu^+ \bar\nu_e$. Generally, these constrain the same operators but are less stringent than the semi-leptonic decay modes. However, we set limits on the scale of all dimension-7 and 9 operators that could lead to a fully leptonic decay, both at tree and loop-level. By using the available experimental limits assuming vector currents we get an estimate of the constraining power of fully leptonic meson decays on different LNV operators. For more precise conclusions, we would like to stress again that dedicated limits for possible scalar current contributions besides the SM would be highly relevant.

When observing a signal pointing to a LNV contribution, the corresponding effective operator would at the same time lead to a contribution to the neutrino mass via loops. With the current limit on the neutrino mass, this similarly implies a corresponding limit on the operator scale. Depending on the loop suppression and the assumed couplings, the limit arising from neutrino masses can be stronger (e.g. operator $\mathcal{O}_{3b}$) or less stringent (e.g. $\mathcal{O}_8$). However, even in the case of a more stringent limit coming from neutrino masses at first glance ($\mathcal{O}_{3b}$), this is subject to the actual UV complete model, as non-trivial cancellation effects or different realizations might change this simplified picture. Hence, it is conservative to not constrain oneself from the beginning by constraints from neutrino masses. To exemplify such a situation, we introduce a simplified leptoquark model with such an non-trivial behaviour.

Finally, we want to stress that the existence of LNV interactions at low scale, such as in rare kaon decays, can have major consequences on the existence of baryogenesis models. LNV interactions realized in observable rare kaon decays would imply strong washout effects of a pre-existing lepton asymmetry and hence could put different leptogenesis models at tension. For a final conclusion, however, two conditions have to be met. First, an equilibration of all flavours need to be guaranteed. For this, an additional sign for lepton flavour violating interactions or complementary signs of LNV in different flavor sectors would be sufficient. Secondly, we would need to confirm that a potential NP contribution to the rare kaon decay is indeed LNV. This would motivate e.g. corresponding searches for LNV interactions at the LHC. The combined implications on mechanisms of neutrino mass generation and the baryon asymmetry strongly motivate the search for LNV processes. Hereby, rare kaon decays are able to probe very high scales of LNV physics.

\acknowledgments
We thank Chandan Hati for useful discussions. K.F. and J.H. were supported by the DFG Emmy Noether Grant No. HA 8555/1-1. F.F.D. acknowledges support from the UK Science and Technology Facilities Council (STFC) via a Consolidated Grant (Reference ST/P00072X/1).